\documentclass[twocolumn]{aastex63}
\usepackage[utf8]{inputenc}
\usepackage{csquotes}
\usepackage{graphicx}	
\usepackage{amsmath, amsthm}	
\usepackage{amssymb}	
\usepackage{multirow}
\usepackage{threeparttable}
\usepackage{appendix}
\usepackage[utf8]{inputenc}
\usepackage[english]{babel}
\usepackage{lipsum}
\usepackage{rotating}

\newcommand{\msun}{\,M$_{\sun}$}
\newcommand{\reff}{$R_{\text{e},\text{NSC}}$} 
\newcommand{\lamR}{${\lambda_R}$}
\newcommand{\vs}{$\left(V/\sigma \right)$}
\newcommand{\lamRe}{${\lambda_R}_\text{e}$}
\newcommand{\vsRe}{$\left(V/\sigma \right)_\text{e}$}
\newcommand{\kms}{\,km\,s$^{-1}$} 
\providecommand{\abs}[1]{\lvert#1\rvert}

\def\deg{\mbox{$^{\circ}$}}

\accepted{July 19, 2021}

\submitjournal{ApJ}

\shorttitle{NSC kinematics}
\shortauthors{Pinna, F. et al.}

\begin{document}

\title{Resolved nuclear kinematics link the formation and growth of nuclear star clusters with the evolution of their early and late-type hosts}

\correspondingauthor{Francesca Pinna}
\email{pinna@mpia.de}

\author[0000-0001-5965-3530]{Francesca Pinna}
\affiliation{Max Planck Institute for Astronomy, 
D-69117 Heidelberg, Germany}

\author[0000-0002-6922-2598]{Nadine Neumayer}
\affiliation{Max Planck Institute for Astronomy, 
D-69117 Heidelberg, Germany}

\author[0000-0003-0248-5470]{Anil Seth}
\affiliation{Department of Physics and Astronomy, University of Utah, 
UT 84112, USA}

\author[0000-0002-6155-7166]{Eric Emsellem}
\affiliation{European Southern Observatory, 
D-85748 Garching bei Muenchen, Germany}

\author[0000-0002-5678-1008]{Dieu D. Nguyen}
\affiliation{National Astronomical Observatory of Japan (NAOJ), National Institute of Natural Sciences (NINS), 2-21-1 Osawa, Mitaka, Tokyo 181-8588, Japan}

\author[0000-0002-5666-7782]{Torsten Böker}
\affiliation{European Space Agency/STScI, 
MD21218 Baltimore, USA}

\author[0000-0002-1283-8420]{Michele Cappellari}
\affiliation{Department of Physics, University of Oxford, OX1 3RH, UK}

\author{Richard M. McDermid}
\affiliation{Research Centre for Astronomy, Astrophysics, and Astrophotonics, Macquarie University, Sidney, 
NSW 2109, Australia}

\author{Karina Voggel}
\affiliation{Universite de Strasbourg, CNRS, Observatoire astronomique de Strasbourg, UMR 7550, 67000 Strasbourg, France}

\author{C. Jakob Walcher}
\affiliation{Leibniz-Institut für Astrophysik Potsdam (AIP)
D-14482 Potsdam}

\begin{abstract}
We present parsec-scale kinematics of eleven nearby galactic nuclei, derived from adaptive-optics assisted integral-field spectroscopy at (near-infrared) CO band-head wavelengths. 
We focus our analysis on the balance between ordered rotation and random motions, which can provide insights into the dominant formation mechanism  of  nuclear  star  clusters  (NSCs).   
We divide our target  sample  into  late-  and  early-type galaxies, and discuss the nuclear kinematics of the two sub-samples, aiming at probing any link between NSC formation and host galaxy evolution. The results suggest that the dominant formation mechanism of NSCs is  indeed  affected  by  the  different  evolutionary  paths  of their  hosts  across  the Hubble  sequence.   More  specifically, nuclear regions  in  late-type  galaxies  are on average more rotation dominated, and the formation of nuclear stellar structures is potentially linked to the presence of gas funnelled to the center.  Early-type galaxies, in contrast, tend to display slower-rotating NSCs with lower ellipticity.  However, some exceptions suggest that in specific cases, early-type hosts can form NSCs in a way similar to spirals.
\end{abstract}


\section{Introduction} \label{sec:intro}
Nuclear star clusters (NSCs) are very compact and bright objects, with average sizes of about 3\,pc \citep[see the review by][]{Neumayer2020}. 
Their sizes, similar to globular clusters (GCs), combined with their higher masses, make them the densest known stellar systems.  Observations of NSCs have proven very challenging since their first detection 
\citep{Redman1937,Babcock1939,Mayall1942}. 
In addition to their small size, they are embedded in the bright central region of their host, 
and require high spatial resolution 
to be disentangled as a different morphological component
 \citep[e.g.,][]{Light1974}. 
While later studies showed that many nearby galaxies of different Hubble types host NSCs \citep[e.g.,][]{Caldwell1983,Kormendy1989,vandenBergh1995,Matthews1997}, only with the resolution of the Hubble Space Telescope (HST) it was possible to find out that NSCs are actually present in
most spiral and early-type galaxies \citep{Carollo1998, Boeker2002,Cote2006}.
We now know that most galaxies of all types between $\sim10^8$ and $10^{10}$\msun\,host NSCs, which are also found in a majority of more massive spiral galaxies \citep{Neumayer2020}. 

Once nucleation was revealed to be a common phenomenon in all kinds of galaxies, the formation of NSCs assumed a greater importance in the context of galaxy evolution. 
One possible formation scenario consists of the orbital decay of stellar clusters towards the galactic center, due to dynamical friction, and their subsequent merging. This was first proposed by \citet{Tremaine1975} and later supported by analytical and numerical studies \citep[e.g.,][]{Capuzzo-Dolcetta1993,Oh2000,Lotz2001}. On the other hand, in-situ star formation after gas infall to the center was proposed as an alternative \citep[e.g.,][]{Mihos1994,Silk1987,Milosavljevic2004,Bekki2015}. 
By and large, NSC formation remains still unclear in spite of the following efforts. 

Scaling relations between NSCs and the underlying galaxy, such as the NSC-to-galaxy mass relation, have clearly pointed to a tight connection of their formation and growth with the evolution of their host \citep[e.g.,][]{GrahamA2003,Ferrarese2006}. 
The nucleation fraction is highest for host stellar masses of about $10^9$\msun, suggesting that the NSC formation mechanisms are most effective in this mass range \citep[e.g.,][]{SanchezJanssen2019}. 
This is true for both early and late-type galaxies, while the decline in nucleation for larger masses is sharp for ellipticals but not clear for massive late types \citep{Neumayer2020}. 
On the other hand, more massive galaxies ($\gtrsim 10^{10}$\msun) seem to host denser nuclear stellar systems \citep{Pechetti2019}. 
Environmental effects may also play a role for early-type galaxies, which appear to be more frequently nucleated towards the central regions of galaxy clusters \citep[e.g.,][]{Ferguson1989,Lim2018}. 

Since different morphological types of galaxies are related to different evolution histories, it is natural to wonder whether they also form their NSCs in a different way. 
Stellar populations can give us important insight on this issue. In late-type galaxies, usually characterised by recent star formation, we observe a mix of nuclear stellar populations with the ubiquity of very young stars \citep[e.g.,][]{Bothun1985,Rossa2006, Seth2006,Walcher2006,Kacharov2018}.
In early-type galaxies it is less clear how common young stars are. Their star formation was often quenched earlier, and their NSCs are often younger than their hosts but do not show signs of on-going star formation \citep[e.g.,][]{Paudel2011,Spengler2017,Kacharov2018}. 
Compared to their early-type hosts, they are more metal rich for host masses above $10^9$\msun, but they are often more metal poor in the low-mass regime \citep[e.g.,][]{Neumayer2020}. 
No similar trend has been observed for late types, but here the picture is complicated by the absence of spectroscopic data in the low-mass regime.

Different observational results have been interpreted as arguments in favour of one or another formation scenario. 
The spatial distribution and the frequency of GCs in nucleated galaxies, as well as the low metallicity of NSCs of low-mass early types, point to the cluster migration scenario \citep[e.g.,][]{Lotz2001, Capuzzo-Dolcetta2009, Lim2018, Fahrion2020}.
On the other hand, the presence of very young populations, extended star formation histories and gradual chemical enrichment in NSCs hosted by late types, strongly support their in-situ formation \citep[e.g.,][]{Seth2006, Walcher2006,Feldmeier2015,Kacharov2018}. 
In addition, the inflow of molecular gas into the central parsecs from larger-scale structures such as bars, disks or rings was mapped in late-type spirals and proposed as the driving mechanism for nuclear starbursts \citep{Schinnerer2003, Schinnerer2006}. 
However, the properties of NSCs in early and late-type galaxies are quite heterogeneous and there is no agreement yet on one dominant formation mechanism. Recent studies point to different origins for the NSCs in different galaxies, while more than one scenario could have played a role in the central region of one galaxy \citep[e.g.,][]{Lyubenova2013,Guillard2016,Kacharov2018}. 

Stellar kinematics, which contain the footprint of the dynamical history of galactic structures, can provide additional constraints for  the origin of NSCs. 
Kinematic maps provided by integral-field spectroscopy (IFS) have revealed that most NSCs rotate, albeit at different levels  \citep{Seth2008b,Seth2010a, Seth2010b,Lyubenova2013,Nguyen2018,Lyubenova2019,Fahrion2019}.
On the one hand, in-situ star formation from gas with high levels of angular momentum is often invoked to explain strong NSC rotation. 
On the other hand, more random motions and complex kinematic structures are usually associated with star-cluster inspiral  \citep[e.g.,][]{Seth2008b,Seth2010a,Seth2010b,Hartmann2011,Nguyen2018}. 
However, recent simulations have supported the idea that observed levels of rotation can be achieved even by star-cluster merging only \citep{Tsatsi2017,Lyubenova2019}.

In this paper we provide a kinematic analysis of the nuclear regions of eleven galaxies across the Hubble sequence, and use it to assess whether a specific scenario is likely responsible for the NSC formation in early and late-type galaxies.
The paper is organized as follows. 
In Sect.\,\ref{sec:sample} we describe the galaxy sample. 
Section\,\ref{sec:obs} gives the relevant information about the data set. 
In Sect.\,\ref{sec:meth} we describe the methods used to perform the kinematic analysis. The results of this are presented in Sect.\,\ref{sec:res} and discussed in Sect.\,\ref{sub:form}. 
We sum up our conclusions in Sect.\,\ref{sec:concl}. 
A more detailed description of the individual host galaxies and their nuclear kinematics, as well as a short discussion on the formation of their NSCs, are included in Appendix\,\ref{app:ind_gal}.

\begin{figure}
\scalebox{0.42}
{\includegraphics[scale=1,page=6]{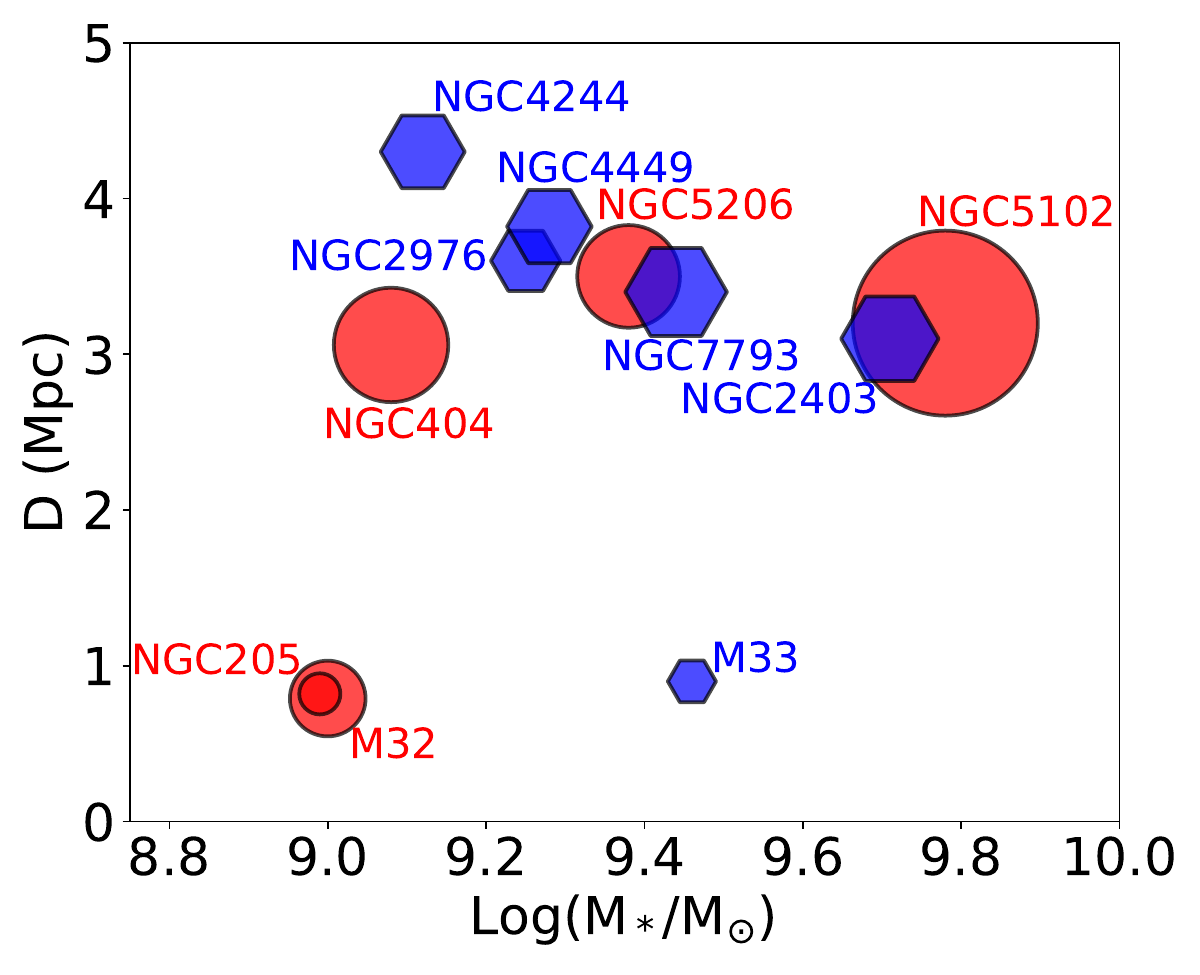}}
\caption{Distribution of the galaxy sample in the distance-stellar-mass plane. Red circles and blue hexagons indicate early-type and late-type galaxies, respectively. The names of the host galaxies are indicated close to the individual points. The size of the symbols is proportional to \reff\,according to Table\,\ref{tab:sample}. 
}
\label{fig:sample_dM}
\end{figure}

\section{The sample} \label{sec:sample}
In this paper, we study the nuclear regions of eleven galaxies with previous NSC detection. 
The sample was selected to be sufficiently nearby (within a distance of 5\,Mpc, see Fig.\,\ref{fig:sample_dM}), to enable parsec-scale resolution, and to contain early and late Hubble types in roughly equal numbers. 
Host stellar masses range from 1 to $6\times10^9$\msun, as illustrated in Fig.\,\ref{fig:sample_dM}. 
Additional relevant properties of the individual host galaxies and their NSCs are indicated in Table\,\ref{tab:sample}. 
The photometric properties of the NSCs (\reff\,and $n_{\rm{NSC}}$, respectively the effective radius and the S\'ersic index) are important for our analysis (see Sect.\,\ref{subsec:vsigma}), and we selected literature data in filters as close as possible in wavelength to our near-infrared (NIR) spectroscopic observations. 

Hints of past interactions, such as warps, shells, tidal tails, and complex kinematic and morphological structures, are common across the sample (see Appendix.\,\ref{app:ind_gal}). 
These galaxies  generally contain gas, albeit at different levels of significance. This is true even for the early types, with the exception of M\,32, which probably has been completely stripped of all gas by past strong interactions. 
Some early types, such as NGC\,404, show extended and complex atomic-gas structures.
As expected, our late-type spirals show intense on-going star formation and some of them show signs of gas inflow towards the center (e.g. galactic fountains in M\,33 and NGC\,2403). 
Additional details on the individual galaxies are given in Appendix.\,\ref{app:ind_gal}. 

\begin{table*}
\centering
\caption{Relevant properties of our galaxy sample and their NSCs. Columns from left to right: galaxy name, alternative name, Hubble type, galaxy distance ($D$), galaxy stellar mass ($M_*$), effective (half-light) radius of the NSC (\reff), S\'ersic index of the NSC ($n_{\rm{NSC}}$). For \reff\,and $n_{\rm{NSC}}$ we chose the values obtained from the HST filters with the closest central wavelength to our NIR data, among the ones available in the literature (see table notes).}
\centering
\begin{tabular}{l|l|c|l|l|r|r} 
\hline\hline
\footnotesize
Galaxy name & Alternative name & Hubble type $^{(1)}$& $D$ (Mpc) & 
$M$$_{\ast}$ ($10^9$\,\msun) & \reff\, (pc) & $n_{\rm{NSC}}$\\ 
\hline\hline
M\,32          &  NGC\,221           &  E       & 0.79 $^{(2)}$ & 1.00 $^{(2)}$  & 4.4 $^{(2)}$ & 2.7 $^{(2)}$\\
M\,33          &  NGC\,598          &  Scd  &  0.9 $^{(3)}$  & 2.9  $^{(6)}$  & 1.8 $^{(3)}$ & 3.8 $^{(3)}$\\
NGC\,205         &  M\,110          &  E      &  0.82 $^{(2)}$ & 0.97 $^{(2)}$ & 1.3 $^{(2)}$ & 1.6 $^{(2)}$\\
NGC\,404   &  UGC\,718           &  S0    &  3.06 $^{(4)}$& 1.2  $^{(4)}$ & 10.1 $^{(4)}$ & 2.6 $^{(4)}$\\
NGC\,2403 &  UGC\,3918        & Scd   & 3.1 $^{(3)}$     & 5.1 $^{(7)}$ & 7.3 $^{(3)}$ & 2.1 $^{(3)}$\\
NGC\,2976 &  UGC\,5221        & Sc    &  3.6 $^{(3)}$     & 1.8 $^{(7)}$ & 3.7 $^{(3)}$ & 5.3 $^{(3)}$\\
NGC\,4244 &  UGC\,7322       &  Scd  &  4.3 $^{(3)}$    & 1.3 $^{(8)}$ & $5.0 - 5.9$ $^{(3)}$ & $1.1 - 2.4$ $^{(3)}$\\
NGC\,4449 &  UGC\,7592       &  Irr     &  3.82 $^{(5)}$ & 1.9 $^{(8)}$ & 5.5 $^{(9)}$ & \\
NGC\,5102  &  ESO\,382-50   & S0     &  3.2 $^{(2)}$    & 6.0 $^{(2)}$ &  26.3 $^{(2)}$ & $0.8 - 3.1$ $^{(2)}$\\
NGC\,5206  &   ESO\,220-18  & S0    &  3.5 $^{(2)}$     & 2.4 $^{(2)}$  &  8.1 $^{(2)}$ & $0.8 - 2.3$ $^{(2)}$
\\
NGC\,7793  &   ESO\,349-12  & Sd    &  3.4 $^{(3)}$     & 2.8 $^{(7)}$  &  7.9 $^{(3)}$ & 3.3 $^{(3)}$\\
\hline
\hline
\end{tabular}
\label{tab:sample}
\begin{tablenotes}
\item {\footnotesize Notes. (1) from NASA/IPAC Extragalactic Database (NED). 
NGC\,205 and NGC\,5206 are classified as dwarf ellipticals (dE) in other references \citep[e.g.,][]{Caldwell1983, Zinnecker1986}.
(2) \citet{Nguyen2018}. \reff\,and $n_{\rm{NSC}}$ were obtained fitting HST images in the following filters: F814W for M\,32, NGC\,205 and NGC\,5206, F547M for NGC\,5102. For NGC\,5206 and NGC\,5102, whose NSCs were fitted with two S\'ersic components, we adopted the integrated \reff. Since $n_{\rm{NSC}}$ is not available for the integrated NSC, we indicate both indices; 
(3) \citet{Carson2015}. \reff\,and $n_{\rm{NSC}}$ were obtained fitting HST images obtained with the F153M filter. We show, for NGC\,4244, the \reff\,and $n_{\rm{NSC}}$ of the two S\'ersic components used to fit the NSC. 
(4) \citet{Seth2010b}. NGC\,404's total mass was estimated using the bulge-to-total luminosity and the bulge mass (Table\,1 in \citealt{Seth2010b}), and assuming a constant mass-to-light ratio. 
(5) \citet{Annibali2008}.
(6) \citet{McConnachie2012}.
(7) \citet{deBlok2008}. 
(8) \citet{Sheth2010}. 
(9) \citet{Georgiev2014}.
}
\end{tablenotes}
\end{table*}


\section{Observations and data reduction} \label{sec:obs}
In this work we used IFS data in the $K$ band, corrected by adaptive optics (AO), of the inner $\sim 3$\,arcsec of the eleven galaxies in our sample. 
We took advantage of facilities in both hemispheres that provide similar capabilities, namely Gemini North in Hawaii, and the Very Large Telescope (VLT) in Chile. Here we provide a summary of the two data sets, and how they were processed prior to the kinematic analysis.

\subsection{NIFS data}\label{sub:nifs}
We reanalyzed in a consistent way previously-published data from the Near-infrared Integral Field Spectrograph (NIFS) \citep{McGregor2003} for M\,32 \citep{Seth2010a}, NGC\,205 \citep{Nguyen2018}, NGC\,404 \citep{Seth2010b}, and NGC\,4244 \citep{Seth2008b}. 
In addition, we present new NIFS data for M\,33, NGC\,2403, NGC\,2976 and NGC\,4449. 
NIFS has a field of view (FoV) of 3\,arcsec\,$\times$\,3\,arcsec and a spectral range between 0.95 and 2.4\,$\mu$m. It is mounted on the Gemini North eight-meter telescope and offers an AO mode thanks to the ALTAIR (ALTtitude conjugate Adaptive optics for the InfraRed) facility \citep{Boccas2006}. 
Laser-guide-star (LGS) AO was used for all galaxies except M\,32, for which a natural guide star (NGS) was used.  

Data reduction was performed using pipelines based on the Gemini version 1.9 IRAF packages, as described in \citet{Seth2008b,Seth2010b}. 
The final cubes were created from several dithered cubes, that were previously telluric corrected and sky subtracted, 
using dedicated exposures collected just before or after observing the science object. 
For the merging of the dithered cubes into the final one, spaxels were rebinned to a size of 0.05\,arcsec\,$\times$\,0.05\,arcsec.
Detailed description of the observations and the data-reduction procedure can be found in \citet{Seth2008b,Seth2010b}, \citet{Seth2010a}, and  \citet{Nguyen2018}. 

The point-spread function (PSF) is described, for the previously-published data, in the respective papers, cited at the beginning of this subsection. 
A two-step approach was used to fit the PSF. A first estimate came from images created from the telluric calibrator. Then, a convolution of a high-resolution HST image with this PSF allowed to refine the fit, by the comparison of these convolved images with the continuum images from NIFS observations. 
The best PSF choice was a double component in most cases, an inner Gaussian and an outer Moffat, while it was a simple Gaussian for NGC\,4244 (see individual references for the details). 
We fitted in this work the PSF of M\,33, NGC\,2403 and NGC\,2976, following the same approach but using a double Gaussian, that gave the best fit. 
The HST images used for the PSF fits were taken with the following two filters: F1042M for M\,32 and M\,33, and  F814W for NGC\,205, NGC\,404, NGC\,2403, NGC\,2976 and NGC\,4244.
Parameters and references of all fits are indicated in Appendix\,\ref{app:PSF}. 
We do not have any PSF measurement for NGC\,4449. The complexity of its central region made it impossible to derive a PSF for our data cubes from optical images. 

With a spectral resolution of $R\sim 5000$, NIFS has an approximately Gaussian line-spread function (LSF), with an average full-width at half maximum (FWHM) of 
$\sim$4.36\,\AA\,\citep{Seth2010b}. 
However, variations in the FWHM of $\sim10$\% were found across the FoV. 
For this reason, the LSF was measured in each spaxel with the help of sky lines (as done e.g. by \citealt{Seth2010b}), in order to achieve more accurate kinematic measurements (see Sect.\,\ref{sec:ppxf}). 

The data cube of NGC\,205 used in this work was star subtracted by \citet{Nguyen2018} to minimize the impact of individual bright stars on the integrated kinematics. 
Their light was subtracted from the data cubes using PampelMuse \citep{Kamann2018b}, as explained in detail in \citeauthor{Nguyen2018} (\citeyear{Nguyen2018,Nguyen2019}). 
Following the same procedure, we obtained a star-subtracted cube also for M\,33. 
While M\,32 is located at a similar distance to NGC\,205 and M\,33, its very dense and bright NSC is crowded with individual stars as observed by NIFS. This made it impossible to use the same approach to subtract these individual stars.

\subsection{SINFONI data}
Observations of NGC\,5102 and NGC\,5206, performed with the Spectrograph for INtegral Field Observations in the Near Infrared (SINFONI) \citep{Eisenhauer2003}, were presented by \citet{Nguyen2018}. 
We reanalysed this data in a consistent way and we added to the analysis new SINFONI observations for NGC\,7793. 
SINFONI was mounted on the Unit Telescope (UT) 4 of the VLT and decommissioned in 2019. 
It offered AO-assisted observations \citep{Bonnet2004} and the NGS mode was used for the corrections of our data cubes. 
SINFONI also has a FoV of 3\,arcsec\,$\times$\,3\,arcsec, with a spaxel size of 0.05\,arcsec\,$\times$\,0.1\,arcsec, covering a wavelength range from 1.1 to 2.45\,$\mu$m.
Data reduction was performed using the ESO SINFONI pipeline, as explained in \citet{Neumayer2007} and \citet{Nguyen2018}.
The final cubes were obtained from a combination of several dithered exposures, after each one of them was corrected for the sky and telluric contaminations. 
Before being combined, SINFONI cubes were also corrected for additive residuals from the sky subtraction and velocity offsets between the individual data cubes, as explained in \citet{Nguyen2018}.

The PSF was evaluated, with the same two-step procedure as for NIFS data (Sect.\,\ref{sub:nifs}) but using two Gaussians for the fit, for NGC\,5102 and NGC\,5206 by \citet{Nguyen2018}. F560W and F814W HST images were used respectively for the two galaxies. We derived the PSF for NGC\,7793 following the same approach using an F153M-band HST image. 
PSF parameters are shown in Appendix\,\ref{app:PSF}. 
SINFONI has a spectral resolution of $R\sim 4000$. 
Since its LSF does not fit a Gaussian function, its actual shape was measured by \citet{Nguyen2018} for NGC\,5102 and NGC\,5206, and in this work for NGC\,7793. The same approach was used for the three galaxies, using OH sky lines, that can be approximated to delta functions. Therefore, the shape of these lines after reaching the detector is a good approximation of the LSF. In fact, the shape of one of these lines when observed with SINFONI appears to be more centrally peaked than a Gaussian and shows broader wings. 
In addition, since the SINFONI LSF varies across columns in the detector, this procedure was followed for each row in the sky frames. 
A median spectral resolution of 6.32\,\AA\,was measured for NGC\,7793.

\section{Methods for the kinematic-analysis} \label{sec:meth}
We extract the nuclear stellar kinematics from NIFS and SINFONI data following a similar approach to \citet{Seth2008b,Seth2010b},  \citet{Seth2010a}, and \citet{Nguyen2018}. 
They previously presented stellar-kinematic maps of the six published data cubes (see Sect.\,\ref{sec:obs}). 
For the sake of completeness and uniformity, we refitted the full sample using the set up described in this section. 
We used the range between 2.285 and 2.390\,$\mu$m, which includes the CO band-head absorption. 
This is the strongest absorption feature in galaxy spectra between 1 and 3\,$\mu$m. It is sharp and deep enough to be very sensitive to stellar kinematics. 
In this region of the spectrum the effects of dust are minimized and there is no emission from the sky \citep[e.g.,][]{Silge2003}.

\subsection{Voronoi binning} \label{sub:vor}
We performed a Voronoi tessellation to spatially bin the reduced and combined NIFS and SINFONI cubes, making use of the VorBin Python package\footnote{\label{note1}
\url{https://pypi.org/project/vorbin/}}. 
The
Voronoi-binning method was described in \citet{Cappellari2003} and consists of a two-dimensional adaptive spatial binning to a minimum signal-to-noise ratio (S/N) around a target value. 
For NIFS data cubes, a variance cube was propagated through the data reduction process. 
The S/N of each spaxel was calculated as the ratio between the mean measured flux (S) and the square root of the mean variance spectrum (N) in the selected wavelength range ($2.285-2.390$\,$\mu$m). 
For SINFONI data, with no available variance spectra, we used the mean measured flux (S) and the standard deviation (N) along the spectral direction, this time in the wavelength range between $2.21-2.26$\,$\mu$m, with no significant absorption features. 
Prior to the Voronoi binning, we set a spaxel S/N threshold of 1 (in the wavelength range used for the analysis), so that all spectra with noise larger than the signal were not binned and discarded from the analysis. 

We initially binned all galaxies to a target S/N of 25 in our wavelength range.
Aiming at a compromise between a spatial resolution good enough to see structures and an acceptable quality of the spectra, we rebinned the nuclei with lower surface brightness (NGC\,2403, NGC\,2976, NGC\,4244 and NGC\,4449) to a target S/N of 15. 
We checked that decreasing the S/N did not have a negative impact on the spectral fitting and on the obtained kinematic parameters. 
Uncertainties (Sect.\,\ref{sub:MC}) were for S/N\,=\,15 similar or lower than for S/N\,=\,25, in particular in the (central) region of the FoV where the Voronoi bins were smaller. 
Using this lower S/N allows us to recover rotation in some regions where it was blurred by larger bins. 
We still preferred to use S/N\,=\,25 for the rest of galaxies, being more conservative when it comes to the goodness of the spectra especially in the outer regions, but reaching already a good compromise with spatial resolution.
We show in Fig.\,\ref{fig:ppxf_ex} three examples of Voronoi-binned spectra with different S/N. 

\subsection{pPXF}\label{sec:ppxf}
\begin{figure}
\scalebox{0.45}
{\includegraphics[scale=1]{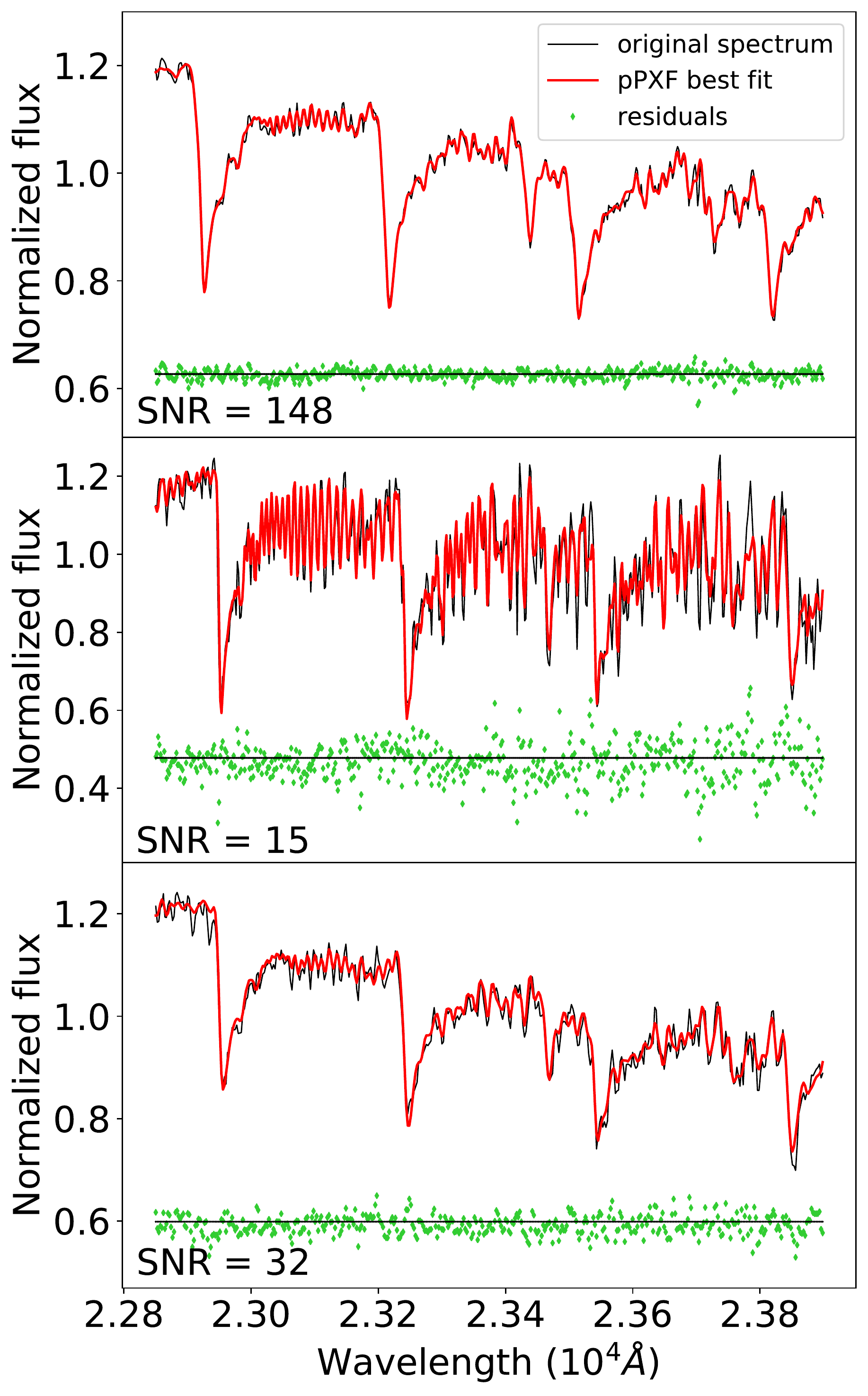}}
\caption{Examples of Voronoi-binned spectra and their pPXF best fits, for different cases. \textit{Top:} M\,32 NIFS spectrum with mean S/N of 148; \textit{middle:} NGC\,4449 NIFS spectrum, S/N\,$ \sim 15$; \textit{bottom:} NGC\,7793 SINFONI spectrum, S/N\,$\sim 32$. Note that the lower values of $\sigma$ in NGC\,4449 (see Fig.\,\ref{fig:ngc4449}) make a much larger number of absorption features visible in the middle spectrum than in the top and bottom ones. 
The observed spectrum is plotted in black, the best fit in red and the residuals of the fit in green. The latter were calculated as the difference between the black and red spectra and arbitrarily shifted to fit in the plot.
}
\label{fig:ppxf_ex}
\end{figure}

We derived the kinematics of the eleven nuclei in our sample, using the Python implementation of the Penalized Pixel-Fitting (pPXF) method\footnote{\label{note2}
\url{https://pypi.org/project/ppxf/}} \citep{Cappellari2004,Cappellari2017}.
This method selects and combines templates from a stellar library and convolves them with a line-of-sight velocity distribution (LOSVD) as a Gauss-Hermite series, to fit the observed spectra. 
We included in the fit the first four Gauss-Hermite moments and used an additive polynomial of fourth order. 
The \enquote{bias} pPXF parameter penalizes the higher moments of the LOSVD, when they are poorly constrained by the data. 
In that case, the LOSVD tends to a Gaussian and the higher moments are biased toward zero. 
We used the default pPXF bias $0.7\sqrt{500/N_{goodpixels}}$, where $N_{goodpixels}\lesssim 500$ is the number of spectral pixels left after masking the bad ones. This number, and then the bias-parameter value, slightly vary depending on the specific spectrum. 
The templates 
were convolved, to match the resolution of the observations, with the LSF measured for the specific Voronoi bin (see Sect.\,\ref{sec:obs}). 
This method provides us maps of mean velocity $V$, velocity dispersion $\sigma$, skewness $h_3$ and kurtosis $h_4$. 
We show in Fig.\,\ref{fig:ppxf_ex} three different examples of pPXF fits. They include spectra from both instruments and with different S/N. 

\subsection{Stellar Library}\label{sec:library}
We initially fitted all galaxies with two different spectral libraries: the second data release (DR2) of the X-Shooter Library (XSL) \citep{Gonneau2020} and the high-resolution spectra of cool stars from \citet{Wallace1996} (hereafter W\&H). 
The XSL was designed to cover most of the Hertzsprung-Russell diagram and includes in total 813 spectra of 666 stars. It covers a large range of spectral types and chemical compositions. Repeated observations were performed for luminous cool stars that were expected to vary in time. 
The spectra cover three contiguous wavelength segments between $\sim 300$\,nm and $\sim 2.45$\,$\mu$m, at a spectral resolution $R \sim 10\,000$.
The W\&H library, instead, provides eight suitable spectra of supergiant, giant, and main-sequence G to M-type stars, of different luminosity classes, at $R \sim 45\,000$. 
This latter set of stars was used for the kinematic analysis by \citet{Seth2010a}, \citet{Seth2010b} and \citet{Nguyen2018}. 

The results obtained with the two spectral libraries were consistent between each other within the error estimates (Sect.\,\ref{sub:MC}). Therefore, the choice of the templates had to be based on the balance between spectral resolution, much higher in W\&H, and number of stars, much higher for the XSL. 
With slightly lower uncertainties in the results, we finally decided to adopt results from the XSL, based on the following considerations. 
According to \citet{Silge2003}, the importance of using a library with a large number of stars resides in the available large variety of CO-band equivalent widths. 
Stars in the XSL were selected to map the widest possible ranges of stellar parameters such as effective temperature and surface gravity, closely related to the equivalent widths 
\citep{Gonneau2020}. 
Although W\&H templates span a large range of equivalent widths in the CO bands (See Table 5 in \citealt{Wallace1996}), 
the range covered by XSL is much more extended and better sampled. 

All spectra in the DR2 of the XSL were corrected for instrument transmission, telluric absorption, radial velocities and were provided in the rest frame (in air wavelengths) \citep{Gonneau2020}. 
From the entire XSL we picked only the spectra whose flux was corrected for slit losses (due to the narrow slit), since a non-corrected slope of the spectra might affect the full spectral fitting. 
We decided to include the repeated observations of variable stars, since they represent stars in different variability snapshots. 
We discarded the spectra with potential issues related to wavy continuum, artefacts or poor S/N in the $K$ band, or superposition of other objects  in the line of sight (LOS) (as from Table\,B.1 in \citealt{Gonneau2020}).
We finally gathered a set of 689 templates.

\subsection{Uncertainties from Monte Carlo simulations}\label{sub:MC}
We estimated uncertainties for each of our kinematic parameters, by means of Monte Carlo (MC) simulations, as suggested by \citet{Cappellari2004}. 
We performed 1000 realizations for each one of the Voronoi bins. In each realization, we perturbed the observed spectra with wavelength-dependent random noise from a Gaussian distribution, since in real spectra noise is not constant in the spectral direction. The standard deviation of the Gaussian 
was set from the variance spectra for NIFS observations, and from the residual spectrum (see Fig.\, \ref{fig:ppxf_ex}) for SINFONI data cubes (with no available variance spectra). 
We ran pPXF for each one of the 1000 perturbed spectra per Voronoi bin, now setting the bias parameter to 0.1. 
This allows for a more conservative estimate of errors as suggested by \citet{Cappellari2004}. 
\citet{Seth2014} and \citet{Nguyen2018} verified the robustness of errors obtained with this approach. 
Typical values of our uncertainties can be seen in all figures in Appendix\,\ref{app:ind_gal}.

\subsection{Ordered rotation vs Random motions}\label{subsec:vsigma}
The ratio of ordered versus random motions \vs\,in galaxies reveals whether these systems are supported by pressure or by rotation and can provide insights about their formation and evolution. 
The anisotropy diagram, \vs\,as function of the observed ellipticity ($\epsilon$), 
was first introduced by \citet{Illingworth1977} and \citet{Binney1978}, to study whether there was a correlation between rotation and the apparent flattening of galaxies. 
The same diagram was later constructed for the 48 early-type galaxies in the SAURON sample \citep{Cappellari2007}. 
In addition, a new parameter \lamR\,was defined by \citet{Emsellem2007}. It has a similar behaviour as \vs\,but is also sensitive to the spatial distribution of the velocity, and is  a proxy of the projected specific angular momentum. 
\citet{Emsellem2007} introduced a new classification of early-type galaxies, into fast and slow rotators, based on \lamR. 
These same \vs\,and/or \lamR\,quantities were later measured for the larger samples of ATLAS$^{\rm{3D}}$ \citep{Emsellem2011}, of SAMI  \citep{vandeSande2017b} and MaNGA galaxies \citep{Greene2018,GrahamM2018}. 
It was shown by the CALIFA collaboration how galaxies of different morphological types populate different regions of the \vs\,and \lamR\,versus $\epsilon$ diagrams \citep{Falcon-Barroso2019}. A review was given by \citet{Cappellari2016}.

These diagrams have been also used to provide an idea of whether NSCs are dominated by rotation or random motions \citep{Lyubenova2019,Neumayer2020}.
In this work, we use them to investigate whether NSCs hosted by galaxies of different Hubble types lie in different regions of these diagrams in the same way as their host galaxies do. 
\vs\,gives a more intuitive way of quantifying the rotational support of a galaxy or a NSC, because it derives directly from the tensor virial equations \citep{Binney1978,Binney2005}. 
However \vs\,gives relatively strong weights to the central part where the flux is usually at its peak. 
On the other hand \lamR, via its radial weighing (see Eq.\,\ref{eq2}), can distinguish central rotation from more extended one, as well as being less dependent on inclination \citep{Emsellem2007,Cappellari2007}. 
For this reason we include both parameters in our analysis. 
We calculated \vs, \lamR\,and $\epsilon$, as luminosity-weighted averages over the area within the half-light (\reff) elliptical isophote of the NSC, using the following definitions from \citet{Cappellari2007} and \citet{Emsellem2007}:

\begin{equation}
\left(\frac{V}{\sigma}\right)_{\rm{e}} =  \sqrt{\frac{\sum{_{i=1}^N}F_i V{_i^2}}{\sum{_{i=1}^N}F_i \sigma{_i^2}}} \label{eq1}
\end{equation}
\begin{equation}
\lambda_{R\rm{e}} = {\frac{\sum{_{i=1}^N}F_i R_i \abs{V_i}}{\sum{_{i=1}^N}F_i R_i \sqrt{V_i^2 + \sigma{_i^2}}}}\label{eq2}
\end{equation}
\begin{equation}
\epsilon_{\rm{e}} = 1-\sqrt{\frac{\sum{_{i=1}^N}F_i y{_i^2}}{\sum{_{i=1}^N}F_i x{_i^2}}}\label{eq3}
\end{equation}

We applied these equations on a spaxel by spaxel basis, on a total number of spaxels $N$ within the \reff\,elliptical isophote. $V_i$ and $\sigma_i$ are respectively the velocity and the velocity dispersion of the i-th spaxel (athough calculated for the corresponding Voronoi bin, Sect.\,\ref{sec:ppxf}).  $F_i$ is the i-th spaxel integrated flux in the original data cube. $(x_i,y_i)$ are the coordinates of the spaxels, with the origin in the nuclear-kinematic centre and rotated to be aligned respectively with the major and minor axis of the \reff\,elliptical isophote. 
$R_i=\sqrt{x_i^2+y_i^2}$ are the distances of spaxels from the center. 
We include in the calculation only spaxels within the \reff\,ellipse, i.e. the elliptical isophote with an area equal to the \reff\,circle ($\pi R^2_{\text{e,NSC}}$). 
In general, the \vs\,and \lamR\,parameters are only rigorously meaningful for isolated systems and not sub-components of galaxies. In the following, we will assume that NSCs are sufficiently dense, hence contrasted above the background and fully dominating the local gravitational potential and light, that they can be modelled as isolated systems with good approximation \citep[e.g.][]{Hartmann2011}.  This will be further examined in Sect.\,\ref{sub:undgal}.

We used the \reff\,in Table\,\ref{tab:sample}, except for NGC\,5102. 
This nucleus was fitted by \citet{Nguyen2018} with two S\'ersic components with effective radii of 0.1 and 2.0\,arcsec. Our NGC\,5102's kinematic maps do not cover the full region out to the integrated \reff\,of 1.6\,arcsec. We chose instead an elliptical isophote of area equal to a circle of a radius of 0.6\,arcsec (Fig.\,\ref{fig:ngc5102}). 
We aimed at finding a compromise between a significant coverage of the FoV and not too large $V$ and $\sigma$ uncertainties. 
The selected radius allows us to include Voronoi bins with uncertainties lower than 17 and 20\kms, respectively for $V$ and $\sigma$, in 90\% of the included spaxel. 
For NGC\,4244 we gave in Table\,\ref{tab:sample} the \reff\,of the two S\'ersic components used to fit the NSC. For the calculations, we used the arithmetic mean of the two values (\reff\,=\,5.43\,pc). 

Uncertainties for \vsRe, \lamRe\, and $\epsilon_{\rm{e}}$ were calculated via MC simulations. We performed here 500 realizations adding random Gaussian noise to the parameters in Eq.\,\ref{eq1}, \ref{eq2} and \ref{eq3}. Each $V_i$ and $\sigma_i$ was perturbed with a noise of the level of their corresponding uncertainties (see Sect.\,\ref{sub:MC} and Fig. from \ref{fig:m32} to \ref{fig:ngc7793}). For $F_i$, we took the mean standard deviation of each spectrum (as used in Sect.\,\ref{sub:MC}).
For $x_i$, $y_i$ and $R_i$, we introduced some noise in the parameters used for their rotation and centering. For the position angle of the \reff\,ellipse, we took the uncertainties from the isophote fitting, while for the position of the kinematic centre we estimated an error of 1 spaxel.

\section{Results} \label{sec:res}
\subsection{Kinematic maps} \label{subsubsec:maps}
For each nucleus in our sample, individual maps of the first four moments of the LOSVD and their uncertainties are shown and discussed in Appendix\,\ref{app:ind_gal}.
Our kinematic results are in excellent agreement with the previously-published maps obtained with the same data sets of M\,32 \citep{Seth2010a}, NGC\,404 \citep{Seth2010b}, NGC\,4244 \citep{Seth2008b}, NGC\,205, NGC\,5102 and NGC\,5206 \citep{Nguyen2018}.
Rotation is observed in all nuclei in our sample, independently of whether the host galaxy is an early or a late type. 
Sometimes this rotation is strong as in M\,32, NGC\,4244 and NGC\,5102, with a maximum velocity between 30 and 60\,\kms, while in others it is not as significant (e.g. NGC\,205 and NGC\,4449 do not reach 10\,\kms). 
Some early-type galaxies display complex kinematic structures in their nuclear regions. This is the case of NGC\,404, whose NSC is made up of two components: the most extended shows clear rotation, while the inner component counter-rotates (see Appendix\,\ref{app:ind_gal} for more details). 
In NGC\,205, we observe an offset between the rotation axis and the minor axis of the nucleus, perhaps suggesting the presence of a merger component.

Regarding the velocity dispersion in the very central region, we observe different behaviours that are clearly related to the mass of the central BH. 
M\,32 and NGC\,5102, with SMBHs of about $10^6$\msun\,\citep[e.g.,][]{Seth2010a,Nguyen2018}, show a strong $\sigma$ peak in their centers. 
In most of our nuclei, with a BH mass of about $10^5$\msun (see Appendix\,\ref{app:ind_gal}), we observe higher $\sigma$ values in the central region but without a defined peak. 
Finally, we have a central drop for low BH masses ($<10^4$\msun) or no BH detections (in M\,33, NGC\,205, and NGC\,2976). 
A similar correlation of the central velocity dispersion with the BH mass was suggested, e.g., by dynamical modelling of GCs and ultra-compact dwarf galaxies (UCDs) \citep{Voggel2018,Aros2020}. On the other hand, galaxies with a SMBH do not show necessarily a central peak of velocity dispersion. The observed LOS $\sigma$ depends as well on BH-unrelated factors, such as the galaxy mass and light distributions, the spatial resolution and the radial anisotropy (see also, e.g., \citealt{McConnell2012}). 

Due to the penalisation against poorly-constrained non-zero values of $h_3$ and $h_4$, introduced by pPXF (Sect.\,\ref{sec:ppxf}), a good determination of these higher moments requires a higher S/N than for $V$ and $\sigma$. We obtain good $h_3$ and $h_4$ maps only for some nuclei in our sample. 
Interestingly, we see a $h_3-V$ anticorrelation in some of the galaxies (M\,32, M\,33, NGC\,404, NGC\,4244, NGC\,5102). This anticorrelation is usually observed in disk-like rotating structures, when superimposed in the LOS on slower or non-rotating components  \citep{vanderMarel1993,Bender1994,Krajnovic2008,Guerou2016,Pinna2019}. 
For the rest of this discussion we focus on the more robust $V$ and $\sigma$ measurements. 

\newpage
\subsection{\vsRe\,and \lamRe\,diagrams and their interpretation} \label{sub:vslam_disc}
\begin{figure}
\scalebox{0.42}
{
\includegraphics[scale=1,page=1,trim={0.5cm 0.7cm 0.5cm 0.7cm}]{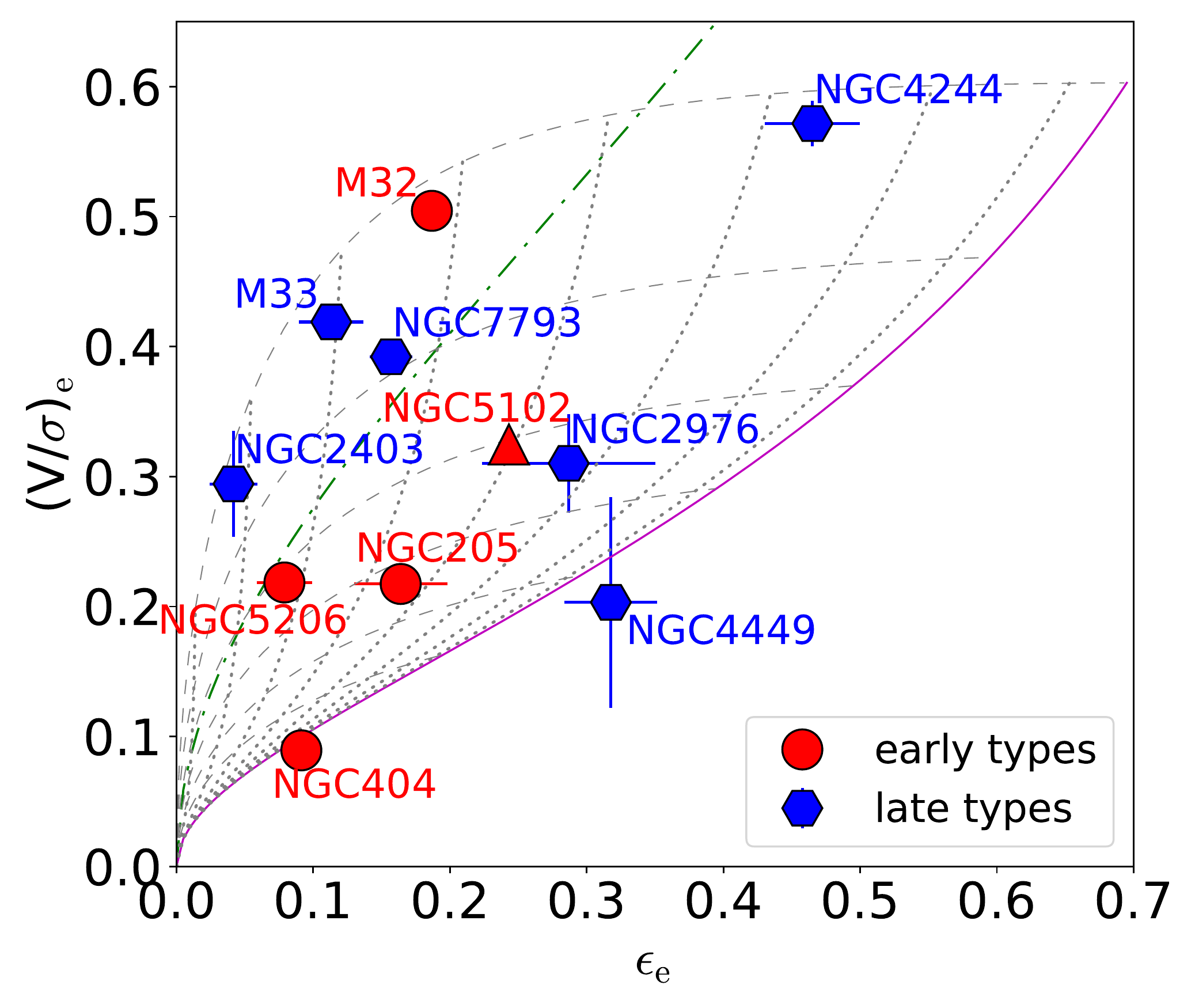}
}
\caption{\vsRe$ - \epsilon_{\rm{e}}$ diagram for the sample of eleven nuclei studied in this work. 
Both \vsRe\,and $\epsilon_{\rm{e}}$ were integrated within the \reff\,ellipse. 
Nuclei hosted by early-type galaxies are indicated with red circles while the ones hosted by late types with blue hexagons. The names of the host galaxies are indicated close to the individual points. 
The upwards triangle indicates a lower limit for NGC\,5102, where the \reff\,ellipse was not fully covered. 
The green dash-dotted and magenta solid lines are indicated as a reference and they refer to edge-on galaxies. They correspond respectively to isotropic oblate rotators  from \citet{Binney2005} and to galaxies with $\delta=0.7\epsilon_{intr}$ 
as from \citet{Cappellari2007}. 
The dotted grey lines show how galaxies with $\delta=0.7\epsilon_{intr}$ move from the magenta line when inclination is decreased. Different lines are separated by steps of 10\deg\,in inclination, from edge on (on the magenta line) to face on. 
Different dashed grey lines correspond to galaxies with different intrinsic ellipticities, from 0.195 (bottom) to 0.695 (top), and with  $\delta=0.7\epsilon_{intr}$. 
}
\label{fig:vsig_sample}
\end{figure}
\begin{figure}
\scalebox{0.42}
{\includegraphics[scale=1,page=1,trim={0.5cm 0.7cm 0.5cm 0.7cm}]{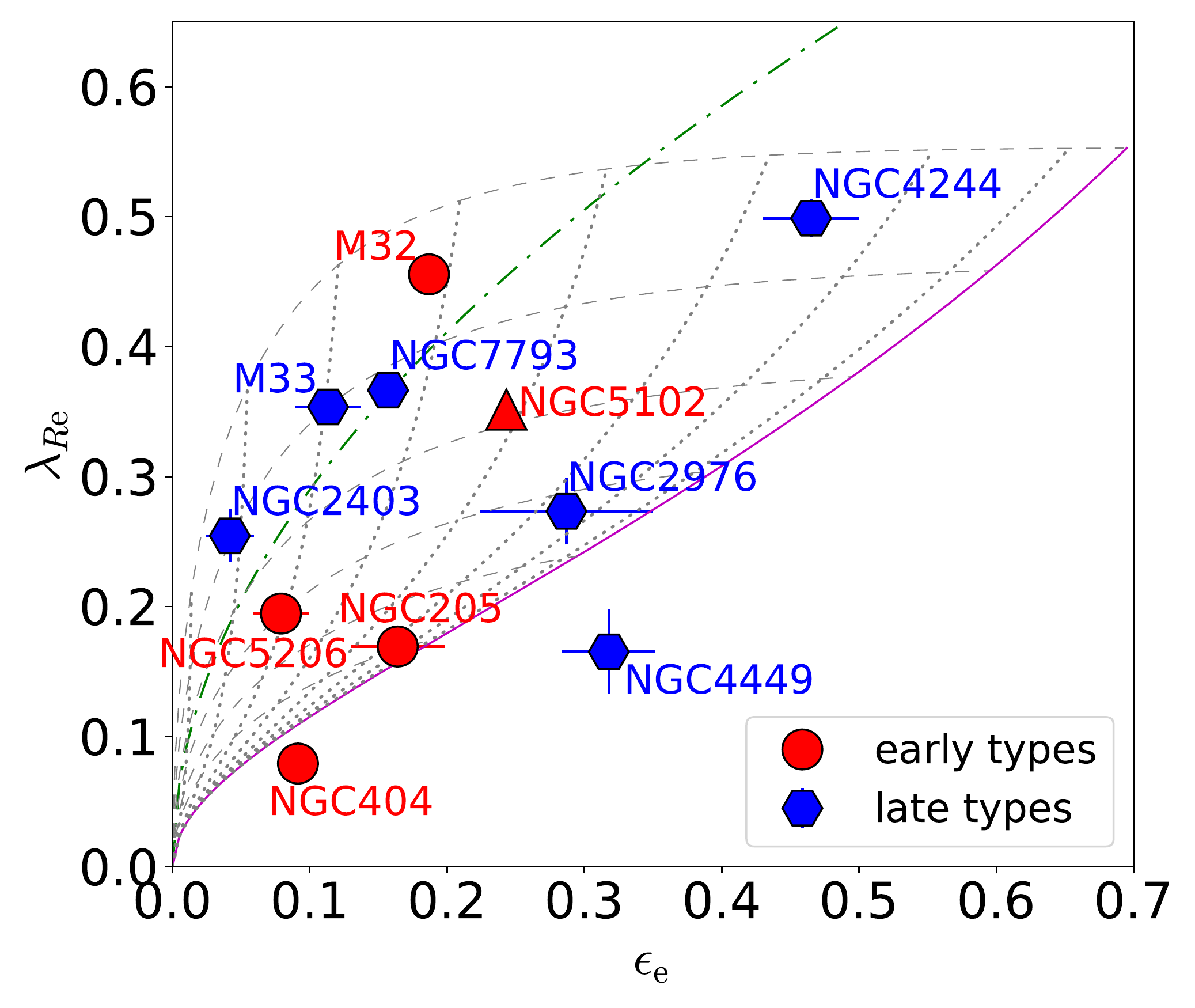}}
\caption{\lamRe$ - \epsilon_{\rm{e}}$ diagram for the sample of eleven nuclei studied in this work. 
Both \lamRe\,and $\epsilon_{\rm{e}}$ were integrated within the \reff\,ellipse. 
Symbols, lines and colors 
are as in Fig.\,\ref{fig:vsig_sample}.
}
\label{fig:lamRe_sample}
\end{figure}

The radial distribution of $\abs{V}/\sigma$ is shown in Appendix\,\ref{app:Vsigma_rad} for each one of the eleven nuclei in our sample. 
Measurements of $\epsilon_{\rm{e}}$, \vsRe\,and \lamRe\,for our full sample are gathered in Table\,\ref{tab:meas} (Appendix\,\ref{app:meas}). 
\vsRe\,and \lamRe\,are plotted as function of $\epsilon_{\rm{e}}$, respectively, in Fig.\,\ref{fig:vsig_sample} and \ref{fig:lamRe_sample}. 
Nuclei of early-type hosts are indicated in red and the ones in late types in blue. 
We show in Fig.\,\ref{fig:vsig_sample} and \ref{fig:lamRe_sample} different lines, which were introduced for galaxies, as a reference to guide the reader's eye. 
The magenta solid line is an approximation for edge-on galaxies with $\delta = 0.7\epsilon_{\rm{intr}}$, where $\delta$ is the velocity anisotropy parameter (\citealt{Binney1987}, Sect.\,4.3) and 
$\epsilon_{\rm{intr}}$ is the intrinsic ellipticity. This line is in general the lower envelope of observed fast-rotating galaxies and was introduced by \citet{Cappellari2007}. 
Lower inclinations move the magenta solid line towards the left side, as indicated by the grey dotted lines. 
Each grey dashed line corresponds to a value of 
$\epsilon_{\rm{intr}}$ and goes, decreasing inclination, from the magenta solid line towards the origin. 
The green dash-dotted line corresponds to \enquote{edge-on isotropic oblate rotators} ($\delta=0$). 
It was introduced by \citet{Binney1978} and updated for integral-field data in \citet{Binney2005}. 
Other roughly parallel lines towards higher $\epsilon_{\rm{e}}$ would correspond to edge-on oblate galaxies with increasing velocity anisotropy $\delta$. 

As expected from the tight relation between \vsRe\,and \lamRe\,\citep{Emsellem2007,Emsellem2011}, the 
sample is distributed in a very similar way in Fig.\,\ref{fig:vsig_sample} and \ref{fig:lamRe_sample}. 
Five of the six nuclei hosted by late-type galaxies are distributed in the upper regions of the diagrams (above 0.25). 
Our five nuclei of early types are distributed all over the ranges of \vsRe\,and \lamRe, with three out of five points below 0.25. 
Therefore, on average, early-type galaxies in our sample show lower nuclear \vsRe\,and \lamRe\,than late types. 
Nuclei in early-type galaxies display also a lower average $\epsilon_{\rm{e}}$ than late types, with values below 0.3 for the full subsample while the highest nuclear  ellipticities correspond to late types.

However, this average trends have to be considered with caution. 
Firstly, the contrast of nuclei with respect to the underlying light components may be lower in early-type galaxies than in late types, having a larger bias on \vsRe\,and \lamRe\,values (see Sect.\,\ref{sub:undgal}). 
This is the case for NGC\,404, the lowest point in the diagrams and the one with the largest contamination from the host galaxy. This nucleus is moreover characterized by other peculiarities. 
The low \vsRe\,and \lamRe\,(measured within \reff) result from the LOS integration of two counter-rotating structures (see Sect.\,\ref{subsubsec:maps} and Appendix\,\ref{app:ind_gal}). 
This nucleus (and the host) has also a low inclination, with a potential impact on the measured $\epsilon_{\rm{e}}$, \vsRe\,and \lamRe (Sect.\,\ref{subsub:incl}). 
If we considered this point as an outlier, early and late-type galaxies would have a similar \vsRe\,and \lamRe\,distribution. 

The three early-type nuclei in the bottom-left corner of the diagrams are characterised by slow observed rotation. 
Not only NGC\,404 but also NGC\,205 display complex kinematic structures. 
The nuclei of NGC\,205 and NGC\,5206, with similar \vsRe\, and \lamRe, show both slow rotation and low ellipticity. 
While also in the nucleus of NGC\,205, we show evidence of different kinematic components, this is not found in NGC\,5206 (Appendix\,\ref{app:ind_gal}). 
Finally, there is one last point in the lower region of Fig.\,\ref{fig:vsig_sample} and \ref{fig:lamRe_sample}. NGC\,4449, a late-type galaxy with low S/N in our nuclear observations and large error bars in the measured kinematic parameters, display \vsRe\, and \lamRe\,values similar to NGC\,205 and NGC\,5206.
However, if we considered this point as an outlier, the trend with late-type galaxies covering the upper region of the diagrams, and early types being located all over the plane, would be much more clear. 

Finally, two of the early-type nuclei lie in the region dominated by late types. 
They are hosted by M\,32 and NGC\,5102 and 
both show a strong disk-like rotation (and a $\sigma$ peak in the center). Moreover, NGC\,5102's plotted values of \vsRe\,and \lamRe\,should be considered lower limits, as discussed later in \ref{subsub:apert}. 
We consider M\,32 and NGC\,5102 two peculiar cases, as it is known from previous studies that they were spirals in the past (see Appendix\,\ref{app:ind_gal}).
Their NSCs probably bring the kinematic footprint of their former late-type host and therefore show similar properties to nuclei of late-type galaxies. 
This will be discussed better in Sect.\,\ref{sub:form_fast}. 

We now discuss some of the impacts that complicate the interpretation of our \vsRe\,and \lamRe\,diagrams.

\subsubsection{Inclination}\label{subsub:incl}
The impact of inclination on the measured values of $V$ (therefore of \vsRe\,and \lamRe) and $\epsilon_{\rm{e}}$, already tested by \citet{Emsellem2007} and \citet{Emsellem2011}, is illustrated by the grey dotted and dashed lines in Fig.\,\ref{fig:vsig_sample} and \ref{fig:lamRe_sample}, referred to galaxies with $\delta=0.7\epsilon_{intr}$. They show that $\epsilon_{\rm{e}}$ is more affected by inclination in the upper half of the diagrams, while \vsRe\,and \lamRe\,are more affected in the leftmost region. 
However, each grey dotted line spans a large range of \vsRe\,and \lamRe (and $\epsilon_{\rm{e}}$), indicating that inclination alone cannot explain the distribution of the global sample in the upper and lower regions of the diagrams. 

Assuming that both NGC\,4244 and its nucleus are seen edge on \citep{Hartmann2011}, inclination might be the main driven for the fact that this nucleus is located in the top-right regions of Fig.\,\ref{fig:vsig_sample} and \ref{fig:lamRe_sample}. 
NGC\,404, the least-inclined galaxy in our sample ($\sim11\deg$, \citealt{delRio2004}), 
hosts a nucleus that is located in the bottom-left corner. 
Apart from the two extreme cases, the rest of the hosts have intermediate-to-high inclinations with no clear correlation with the values of \vsRe\,and \lamRe. 
Nuclei in the bottom region (e.g. the one in NGC\,205, with inclination $\sim 59\deg$, \citealt{Nguyen2018}) are not necessarily the ones with the lowest inclinations. 
The fact that the nucleus hosted by NGC\,5206 ($\sim44\deg$ inclined, \citealt{Nguyen2018}) is located in Fig.\,\ref{fig:vsig_sample} close to the same dashed line as the one in NGC\,5102 ($\sim72\deg$ inclined, \citealt{Nguyen2018}), may suggest that the two galaxies might have similar intrinsic levels of rotations but observed at different inclinations. However, this does not happen in Fig.\,\ref{fig:lamRe_sample}, where these two galaxies lie close to (dashed) lines with different intrinsic ellipticities. 

Some blue hexagons, mostly in the upper half of the diagrams, correspond to relatively low host inclinations. 
An example is M\,33's nucleus, whose trajectory in the \vsRe$ - \epsilon_{\rm{e}}$ diagram if projected from its inclination ($\sim 49\deg$, assumed to be the same as the host galaxy) to the edge-on view was shown by \citet{Hartmann2011} in their Fig.\,5. 
The nuclei of M\,33 and NGC\,2403, very similar galaxies (see Appendix\,\ref{app:ind_gal}), lie relatively close to each other in the diagrams and their offset might be explained at least partially by their different inclination (independently of NGC\,2403's inclination, its nucleus shows almost round isophotes in Fig.\,\ref{fig:ngc2403}).

\subsubsection{Integration aperture}\label{subsub:apert}
The integration aperture has also an impact on measured values of $\epsilon$, \vs\,and \lamR\,\citep[e.g.,][]{Emsellem2007,Emsellem2011}. 
In galaxies, a larger integration radius corresponds in general to a larger \lamR. However, this depends on the specific structure of galaxies, and later-type galaxies with an underestimated $R_{\rm{e}}$ may lie in the same region of the \lamRe\,diagram as galaxies with a larger bulge but with an overestimated $R_{\rm{e}}$ \citep{Harborne2019}. 
Similarly, more disky NSCs with an underestimated \reff\,might lie in the same region as more spherical or slower-rotating NSCs with an overestimated \reff. At the same time, an overestimate of \reff\,would lower the contrast of the NSC with respect to the underlying galaxy, affecting the self-consistency of the measurements (Sect.\,\ref{sub:undgal}).

For NGC\,5102 we used an aperture smaller than the \reff\,ellipse, because the latter was not entirely covered by our data and due to the larger errors in the outer region of our FoV (Sect.\,\ref{subsec:vsigma}). Therefore, we expect a bias towards lower values in our measurements of \vsRe\,and \lamRe. 
Aperture corrections were proposed 
by \citet{vandeSande2017a} to quantify this kind of bias 
for the SAMI and ATLAS$^{\rm{3D}}$ galaxy surveys. 
However, as these corrections are PSF dependent and have only been applied to galaxies, they are likely inappropriate for NSCs. Thus we chose to conservatively treat the non-aperture corrected values of NGC5102 as lower limits and to indicate them with upwards triangles in Fig.\,\ref{fig:vsig_sample}, \ref{fig:lamRe_sample}, \ref{fig:vsig_comp} and \ref{fig:lamRe_comp}. 
\begin{figure}
\scalebox{0.42}
{\includegraphics[scale=1,page=1]{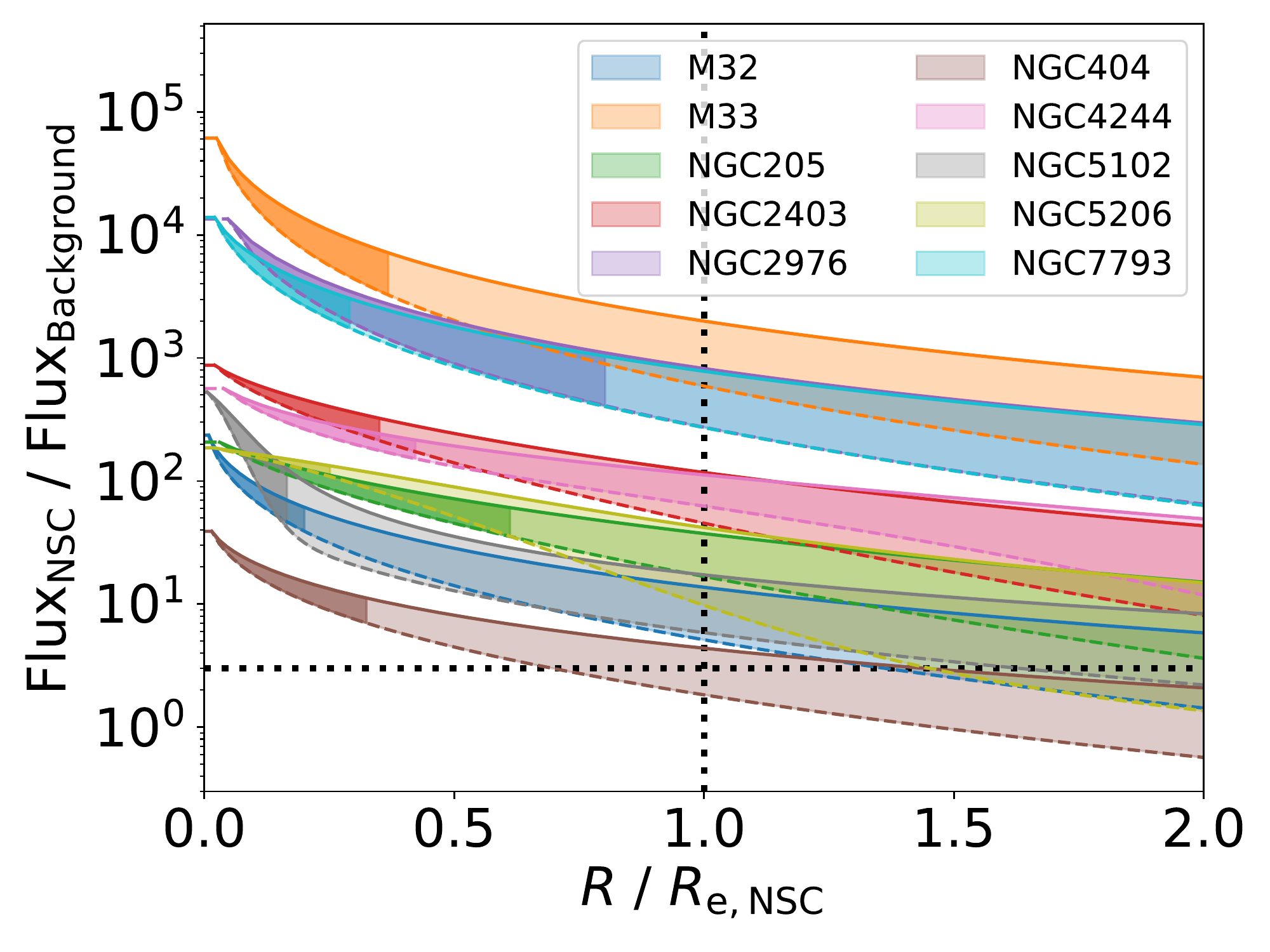}}
\caption{Contrast between the NSCs and the background flux for each galaxy in our sample (except NGC\,4449), illustrated by the radial profiles of the ratio between the surface brightness of the two components. Each galaxy is represented by an individual colour. The shaded areas are limited by an upper (resp. lower) curve corresponding to the ratio of the integrated (resp. local) flux. The darker shaded areas are added to represent the radius of the HWHM, the half-width at half maximum, of the PSF for each galaxy observation. The vertical dotted line shows the 1~\reff\,radius, while the horizontal dotted line is just indicative of a factor of 3 ratio.
}
\label{fig:contrast}
\end{figure}

\subsubsection{PSF effects}
Kinematic measurements are known to be affected by the PSF when this is of the order of the characteristic size of the galaxy \citep{Greene2018,GrahamM2018,vandeSande2017b,Harborne2019,Harborne2020}. Beam smearing and atmospheric seeing result in observed lower rotation velocity and higher velocity dispersions in the central region of the FoV. Therefore, measurements of \vsRe\,and \lamRe\,are in general biased towards lower values. The impact of seeing is larger at scales comparable to the PSF and for galaxies with a higher amount of rotation \citep{Harborne2019,vandeSande2017b,GrahamM2018}. 

\citet{Harborne2020} proposed analytic corrections to estimate the impact of the PSF in observational measurements of \vsRe\,and \lamRe\,in galaxies (Appendix\,\ref{app:PSF}). 
While detailed modelling should be done in order to assess whether these corrections apply to the specific regime of galactic nuclei, 
and they might not be fully appropriate for a quantitative analysis, they provide a qualitative idea of the PSF effect. We have used them to estimate how important this impact is in our measurements. We include the details of the calculations and the results in Appendix\,\ref{app:PSF}. 
The PSF corrections shift all nuclei upwards in the \vsRe\,and \lamRe\,plots (Fig.\,\ref{fig:vsig_corr} and \ref{fig:lamRe_corr} in Appendix\,\ref{app:PSF}). 
This effect is stronger for nuclei hosted by late-type galaxies (an increase of $\sim$60\%) than for those in early types ($\sim$40\%, see also Table\,\ref{tab:meas} in Appendix\,\ref{app:meas}). 
NGC\,205 is an exception, with the largest PSF impact among early-type galaxies, mainly due to the small size of his NSC relatively to the PSF size. 
Nevertheless, the points remain similar in relative position to each other, and thus this correction does not significantly affect our conclusions. 

\subsubsection{The impact of the underlying galaxy}\label{sub:undgal}
Our method is based on a LOS-integrated analysis of kinematics. Since NSCs are embedded in their host galaxy, kinematic measurements in the nuclear region are affected by all other components in the LOS.
In this paper, we assume that the light from the NSC fully dominates the region within its \reff\,and therefore the bias of our measurements due to the underlying galaxy is not significant.

We first tested this by deriving both the local (surface-brightness) and integrated (within an aperture) radial flux profiles of the NSCs relative to the background galaxy. For such an experiment, we used the surface brightness decomposition as available via \citet{Seth2010b},  \citet{Carson2015} and \citet{Nguyen2018}, (as from the rightmost column in Table\,\ref{tab:sample}). In Fig.~\ref{fig:contrast}, we show the ratio of the local and integrated fluxes between the NSC and the background, for all galaxies in our sample except NGC\,4449, with no surface brightness profiles available in the literature. Within 1~\reff, those ratios are almost all above a value of 10, while only NGC\,404 shows a ratio around 3 at that radius. This confirms that the NSCs fully dominate both the local and integrated flux budget within the 1~\reff\,apertures.

The level of contamination on the measured values of \lamRe\,and \vsRe\,depends on many parameters, including the gradient of the potential assumed for the background, the mass ratio between the two components, the compactness of the NSC, the mass-to-light ratio (as both the potential and the luminosity weighing play a role), and obviously the dynamical status of each component (their relative $V$ and $\sigma$ values which spatially vary). 
We performed a simplified calculation taking into account the weighing of each component in the derivation of \lamRe\, (or \vsRe) but assuming a rather constant luminosity for the background. This shows that when these ratios stay above a value of 10, the contamination from the host potential and light is relatively little, at the level of 10 to 15\% (assuming mass follows light within \reff). For NGC\,404, the contamination may be higher and we estimate, again using the same simple assumptions, that it amounts to a maximum of 40\%.

To check such a naive calculation, we further produced mock kinematic models of the central region of three targets in our sample, namely NGC\,404, M\,32 and NGC\,7793, representing the two worst as well as one of the best cases, respectively, in terms of the NSC contrast. We thus directly examined the impact of embedding the NSC into the potential well of an host galaxy, by comparing the resulting measures of \vsRe\,and \lamRe\,in the ideal case of an \enquote{isolated} self-gravitating NSC, and of the same NSC within its late-type host.

We first modelled the mass distribution of the three galaxies via Multi-Gaussian Expansions \citep[hereafter, MGE]{Emsellem1994,Cappellari2002}.
For NGC\,404, we made use of the model published by \cite{Nguyen2017}. For M\,32, we used the MGE model by \cite{Verolme2002} and for NGC\,7793 we built a model based on an image from the 2MASS Large Galaxy Atlas \citep{Jarrett2003}.
We then predicted the projected nuclear kinematics (V, $\sigma$) assuming axisymmetry and a fixed inclination and computed the respective integrated \lamR\, (and \vs) radial profiles. As expected, when adding the underlying galaxy background potential (and light) to the NSC, measurements tend to higher or lower values depending on the dynamical state of these background stars. Again, the impact naturally depends on both the local fraction of light corresponding to the NSC and on its integrated weight within the selected aperture. 

As expected, the effect is negligible for \lamR\,and \vs\,values within radii where the NSC fully dominates with contrast ratios above 10, while it becomes significant for ratios below 5. For both NGC\,7793 and M\,32, the contamination ($\Delta \lambda_R$)
is less than $10\%$ while for NGC\,404 it is just below 20\%. An interesting twist pertaining to the case of NGC\,404 is that the nuclear region includes clear stellar counter-rotation, which, together with its low inclination, tend to make its observed \lamR\,value quite low. The photometric decomposition used in the above-mentioned modelling assumes that the NSC is a mixture of co- and counter-rotating stars, and has a very strong azimuthal anisotropy (lowering its mean stellar velocity). If we were to assume that the NSC is made only of the most central counter-rotating stars, it would maximise its mean velocity while increasing then the relative anisotropy difference between the NSC and the host, and would in turn very significantly emphasise the contamination. 

Overall, we therefore conclude that our \lamRe\,and \vsRe\,are not significantly affected by such a contamination effect, except possibly for NGC\,404, which has a complex kinematic structure \citep{delRio2004, Bouchard2010}, and quite a low inclination.

\subsection{Previous samples in the literature}\label{sub:comp}
We added previously-observed NSCs, GCs and UCDs, with available kinematics in the literature, to the \vsRe$ - \epsilon_{\rm{e}}$ and \lamRe$ - \epsilon_{\rm{e}}$ diagrams. 
In Fig.\,\ref{fig:vsig_comp} and \ref{fig:lamRe_comp}, we compare our results with a sample of six nuclei from \citet{Lyubenova2019}, indicated with red open circles, and with measurements from \citet{Feldmeier2014} for the NSC in our Galaxy, indicated as a blue open upward triangle. 
We also included in the diagrams a sample of GCs in the MW from \citet{Kamann2018a} (indicated as orange open plus symbols), the nucleus of the Sagittarius dwarf galaxy, M\,54, from \citet{AlfaroCuello2020}  (brown open star), and two UCDs, M\,59-UCD\,3 from \citet{Ahn2018} and M\,60-UCD\,1 from \citet{Seth2014} (brown open "X" symbols). 

\subsubsection{Other galactic nuclei} \label{sub:publ_NSCs}
We compare here our sample with 
the six nuclei whose kinematics was studied by \citet{Lyubenova2019}. 
They are hosted by early-type galaxies in the Fornax cluster and thus they  live in a higher-density environment and they are more than five times more distant than our sample (about $\sim20$\,Mpc from us). 
While this leads to a more limited physical resolution than in our sample, we have a similar number of Voronoi bins, within the \reff, to some of our galaxies. 
The galaxies in this sample have also, on average, larger masses than ours. 
\citet{Lyubenova2019} extracted the nuclear kinematics, from AO-assisted SINFONI data, using a very similar approach to ours. 
On the other hand, they estimated $\epsilon_{\rm{e}}$ from isophotal fitting within \reff. 

Similar to the early-type galaxies in our sample, most of the nuclei in their sample are round and do not rotate strongly, as shown in Fig.\,\ref{fig:vsig_comp} and \ref{fig:lamRe_comp}. Four out of six have \vsRe$<0.25$, \lamRe$<0.2$ and $\epsilon_{\rm{e}}< 0.2$ and are actually located in the diagrams close to the lowest points in our sample (especially to NGC\,404). These points strengthen the trend with nuclei in early-type galaxies being located in a lower region of the diagrams with respect to the ones hosted by late types. However, caution is needed regarding this trend since early-type galaxies from \citet{Lyubenova2019} are different objects from galaxies in our sample. As mentioned, they are more massive and live in a much denser environment, which implies accelerated mass assembly and early quenching \citep[e.g.,][]{Fujita1999}.
On the other hand, just as with our sample, there are two exceptions of strongly-rotating nuclei hosted by early-type galaxies, FCC\,47 and FCC\,170. 
These nuclei have peculiar properties that might be the cause for their location in the diagrams. Their origin is discussed in Sect.\,\ref{sub:form}. 

\begin{figure}
\scalebox{0.42}
{\includegraphics[scale=1,page=2,trim={0.5cm 0.7cm 0.5cm 0.7cm}]{ellipt-Vsigma_paper_incllines_ucdsinner_allLSFPSF.pdf}}
\caption{Comparison of our sample of eleven nuclei (filled points, as from Fig.\,\ref{fig:vsig_sample}) with points from previous works (open points), in the \vsRe$ - \epsilon_{\rm{e}}$ diagram. Red open circles correspond to nuclei hosted by early-type galaxies analysed by \citet{Lyubenova2019}. The open blue upwards triangle corresponds to a lower limit for the MW \citep{Feldmeier2014}. Orange open plus symbols correspond to GCs from \citet{Kamann2018a}. M\,54 \citep{AlfaroCuello2020} is indicated with a brown star and the two UCDs M\,59-UCD\,3 \citep{Ahn2018} and M\,60-UCD\,1 \citep{Seth2014} with brown open "X" symbols. 
The green dash-dot and the magenta solid lines are as in Fig.\,\ref{fig:vsig_sample}. 
}
\label{fig:vsig_comp}
\end{figure}

We calculated \vsRe\,and \lamRe\,for the NSC in the MW, using values of $V$ and $\sigma$ from \citet{Feldmeier2014}. Their radial coverage was slightly smaller than \reff, therefore this point should be considered as a lower limit and is indicated as an upwards triangle in Fig.\,\ref{fig:vsig_comp} and \ref{fig:lamRe_comp}. 
The ellipticity (0.29) was taken from \citet{Schoedel2014}. 
The NSC in the MW lies in the upper region of the diagrams, providing additional evidence that late-type nuclei are, on average, more rotation dominated than early-type nuclei.
While the MW is more massive than galaxies in our sample, with a stellar mass of $\sim 6\times 10^{10}$\msun\,\citep{Licquia2015}, and hosts a SMBH of $\sim4\times 10^6$\msun\,\citep{Gillessen2017}, 
the mass of its NSC is similar to the most massive NSCs in our sample (hosted by M\,32 and NGC\,5102, see Appendix\,\ref{app:ind_gal}). With a quite typical \reff$\sim 4.2$\,pc, the MW NSC has a relatively high mass of $1.4-2.5\times 10^7$\msun\,\citep{Schoedel2014,Feldmeier2014,Feldmeier2017b}. 
In Fig.\,\ref{fig:vsig_comp} and \ref{fig:lamRe_comp}, it is located relatively close to massive nuclei in our sample. 
We discuss in Sect.\,\ref{sub:form} on the potential formation scenario of the NSC in our Galaxy. 
\begin{figure}
\scalebox{0.42}
{\includegraphics[scale=1,page=2,trim={0.5cm 0.7cm 0.5cm 0.7cm}]{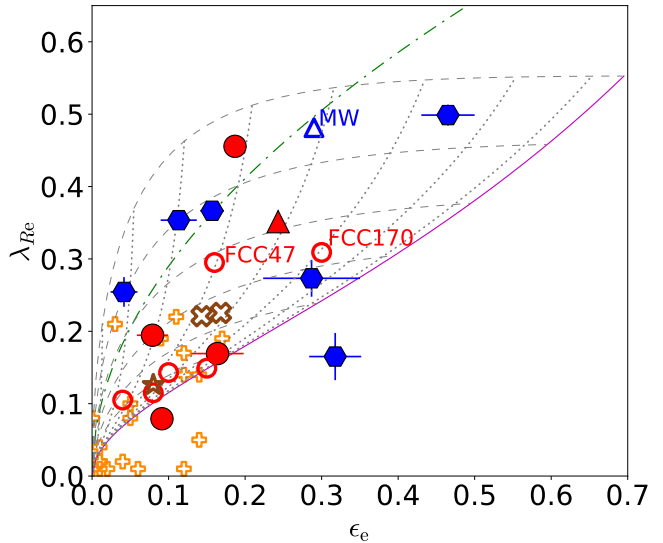}}
\caption{Comparison of our sample of eleven nuclei with points from previous works, in the \lamRe$ - \epsilon_{\rm{e}}$ diagram. 
Symbols, lines and colors are as in Fig.\,\ref{fig:vsig_comp}.
}
\label{fig:lamRe_comp}
\end{figure}

\subsubsection{Globular clusters and ultra-compact dwarfs}
GCs are connected to NSCs in different ways, although they are systems with different intrinsic properties and they reside in different regions of the potential well of galaxies. 
The relationship between NSCs and GCs is rather complex: they could either form in a similar way, NSCs could form from GC accretion, and some GCs are stripped galaxy nuclei. 
To facilitate a comparison, we included in our discussion a sample of 21 MW GCs with published \vsRe\,and \lamRe\,(\citealt{Kamann2018a}, ellipticities from \citealt{Harris1996}). 
The full GC sample in Fig.\,\ref{fig:vsig_comp} and \ref{fig:lamRe_comp} lies in the region where most nuclei hosted by early-type galaxies are concentrated, perhaps suggesting a potential evolutionary connection with these NSCs (see Sect.\,\ref{sub:form}). 

We have added to our diagrams M\,54, an object straddled between NSCs and GCs and with available kinematics. It is the second most massive GC of the MW, and the stripped nucleus of the disrupted Sagittarius dwarf spheroidal galaxy \citep[][see discussion in Sect.~\ref{sub:gc_ucd}]{Bellazzini2008,Mucciarelli2017,AlfaroCuello2019}. 
We calculated \vsRe\,and \lamRe\,for M\,54 from the kinematics published by \citet{AlfaroCuello2020}, approximating a constant flux per spatial bin within the effective radius. This point is located, in Fig.\,\ref{fig:vsig_comp} and \ref{fig:lamRe_comp}, in the region dominated by NSCs in early-type galaxies and by the faster-rotating GCs.

Finally, we included in our analysis the two UCDs M\,59-UCD\,3 and M\,60-UCD\,1, for which the kinematics is available respectively from \citet{Ahn2018} and \citet{Seth2014}. We added these two points to Fig.\,\ref{fig:vsig_comp} and \ref{fig:lamRe_comp}, integrating within the effective radius of the inner morphological components of these UCDs, which are thought to be the NSCs of the progenitor galaxies. These points are actually located, in our diagrams, on the top of the early-type dominated region, where the fastest rotating GCs are found. 

\section{Discussion}\label{sub:form}
While some amount of rotation was observed in all nuclei in our sample, they are distributed in different regions of the \vsRe\,and \lamRe\,diagrams. 
Overall, our results suggest that most NSCs in late-type galaxies are rapidly rotating, while those in early-type galaxies have on average less rotation. 
In this section we discuss how these observations translate to constraints on NSC formation. 
We first consider the interpretation of galaxy kinematics in the context of their formation to see if analogous formation mechanisms may be at work in NSCs.
\citet{Emsellem2007} suggested that gas and mergers play a key role in the formation and evolution of respectively fast and slow-rotating early-type galaxies. 
\citet{Greene2018} found that an important fraction of high-mass early-type galaxies have high angular momentum, and they differ from slow rotators in their (larger) amount of ionized gas. 
The importance of gas, either provided via simple accretion or wet mergers, to keep and/or increase the angular momentum of a galaxy, was also shown in simulations \citep[e.g.,][]{Lagos2018,Walo2020}.

If we qualitatively extrapolate this interpretation to galactic nuclei, their location in the \vsRe\,and \lamRe\,diagrams might be determined by the balance between the two most invoked NSC formation mechanisms, in-situ formation from gas inflow and inspiralling of star clusters. While they might have both played a role in the formation of the same NSC, we use our sample to test a scenario where a more dominant contribution of in-situ star formation, with respect to star-cluster infall, would correspond to larger values of \vsRe\,and \lamRe. 
However, we have to be cautious to avoid too simplistic interpretations. The full picture needs to include other possibilities such as the accretion of gas-rich clusters \citep{Guillard2016} or the accretion of clusters with high angular momentum as the ones in \citet{Lahen2020}. 

Appendix\,\ref{app:ind_gal} contains, for each individual nucleus in our sample, the detailed description of the kinematics as from our results, as well as 
a review of previous studies regarding both the NSCs and host galaxies. In this broader context, a short discussion on the potential formation scenario for each NSC is also provided. 
We therefore summarize and further discuss here the information contained in Appendix\,\ref{app:ind_gal}. 

\subsection{The formation of higher-angular-momentum NSCs}\label{sub:form_fast}
In our sample, galactic nuclei in late-type galaxies show on average higher values of \vsRe, \lamRe\,and $\epsilon_{\rm{e}}$. 
This is in agreement with in-situ formation playing a leading role in the formation of these NSCs, hosted by a type of galaxies that is in general characterized by on-going star formation. 

As recently suggested for early-type galaxies by \citet{Fahrion2021}, gas accretion and subsequent in-situ formation might be needed for the growth of massive NSCs ($\gtrsim 10^7$\,\msun). However, this may happen in all types of galaxies (see also \citealt{Neumayer2020}).
The four most massive NSCs in our sample, with masses larger than $10^7$\,\msun\,(see Appendix\,\ref{app:ind_gal}), are hosted by M\,32, NGC\,5102, NGC\,4244, and NGC\,7793. We note that all of them are represented by points located in the upper region of the diagrams in Fig.\,\ref{fig:vsig_sample} and \ref{fig:lamRe_sample}. 
This is consistent with a dominant contribution from highly-rotating and/or dynamically cooler stars formed in situ after gas inflow. 
In previous studies, this formation channel was in fact proposed as potentially dominant in the NSCs hosted by M\,32 and NGC\,5102 \citep[e.g.,][]{Seth2010a,Nguyen2018}, but necessary to explain observations also in NGC\,4244 and  NGC\,7793 (e.g., \citealt{Hartmann2011,Kacharov2018}, see also Appendix\,\ref{app:ind_gal} for the individual cases). 
In-situ formation might be therefore the dominant mechanism in late-type NSCs and massive, rapidly-rotating NSCs in early-type galaxies.

\subsection{The formation of lower-angular-momentum NSCs}\label{sub:form_slow}
In the bottom-left region of Fig.\,\ref{fig:vsig_sample} and \ref{fig:lamRe_sample} we have three of our nuclei, hosted by early-type galaxies. 
The nucleus of NGC\,404, represented by the lowest point in the diagrams, shows two counter-rotating structures, while 
the one in NGC\,205 displays an offset between the rotation axis and the minor axis. 
In both cases, the complex kinematics suggest a formation in different episodes with one or more mergers playing a major role. 
As mentioned in Appendix\,\ref{app:ind_gal}, these mergers would have allowed the accretion of star clusters and/or triggered star formation bursts. 
On the other hand, no kinematic-decoupled components are detected in the nucleus of NGC\,5206. However, we know from the photometric decomposition \citep{Nguyen2018} that this NSC is also made up of different components: an inner one and a more extended one. This NSC shows a significant contribution from a continuous in-situ star formation (\citealt{Kacharov2018}, see also Appendix\,\ref{app:ind_gal}). 

In summary, while some amount of nuclear in-situ star formation might be ubiquitous for all kinds of galaxies, this mechanism seems to dominate in NSCs of late-type galaxies. On the other hand, the picture is not as clear for early-type galaxies. The latter are characterized by more complex structures and a large variety of different cases. 

\subsection{The formation of NSCs in the context of their host galaxies}\label{sub:form_hosts}
In our sample, the ubiquity of rotation, together with the signs of past interactions, is very often in conjunction with peculiar gas structures. 
As from Appendix\,\ref{app:ind_gal}, the presence of some amount of gas was previously shown in all our host galaxies except M\,32, probably because of its dramatic past stripping processes related to its interaction with M\,31 \citep[e.g.,][]{Dierickx2014}. 
This suggests that gas accretion might play a key role in the formation and growth of NSCs, leading either to nuclear in-situ star formation or to clustered star formation in the galactic disk (outside the nucleus) followed by a migration towards the center. 

Very often, the gas structures observed in our sample were associated with past interactions. The latter might be fundamental for nucleation, as they can provide not only external gas but also star clusters from the accreted satellites. In addition, they can trigger in-situ star formation. 
Evidence for gas inflow towards the galactic center, in the shape of filaments, bars or galactic fountains, was previously provided for four out of six late-type galaxies in our sample (M\,33, NGC\,2403, NGC\,2976, NGC\,4449, see Appendix\,\ref{app:ind_gal} and references therein). In addition, signs of very recent or ongoing in-situ star formation 
suggest that this plays in general a major role in assembling NSCs of late-type galaxies. 

The strong rotation in M\,32's and NGC\,5102's nuclei, the two points associated to early-type galaxies that are located in the upper region of the \vsRe\,and \lamRe\,diagrams, may be related to the peculiar past of their hosts as spiral galaxies. 
M\,32 is considered the stripped remnant of a more massive spiral 
\citep[e.g.,][]{Dierickx2014,dSouza2018}. NGC\,5102, the most massive galaxy in our sample, is a peculiar lenticular sharing properties with late-type galaxies, such as its extended atomic gas component. It would have consumed most of its gas during past intense star formation and merger episodes, becoming a lenticular \citep[e.g.,][]{Davidge2008b}. 
These two NSCs probably kept the kinematic properties of their disky progenitors and/or they were formed 
via in-situ star formation triggered by the interactions which led to their galaxy-type transition (see Appendix\,\ref{app:ind_gal} for more details and references). 
To sum up, all nuclei in the upper region of Fig.\,\ref{fig:vsig_sample} and \ref{fig:lamRe_sample} (all hosted by late-type galaxies plus two early types) have shown in previous works signs of gas inflow and/or in-situ star formation. 
This is in agreement with a major role of in-situ formation in those nuclei that are segregated in the upper region of the \vsRe\,and \lamRe\,diagrams.

On the other hand, intense merger histories may lead to lower values of \vsRe\,and \lamRe,  when associated to complex nuclear structures. 
The complexity found in the nucleus of NGC\,404, the lowest point in the diagrams, extends to the large scales of the host galaxy, and is consistent with the merger NSC origin suggested by our results (Appendix\,\ref{app:ind_gal}). 
This lenticular galaxy displays a peculiar gas structure made up of different kinematically decoupled components, while the complex stellar populations also result from the combination of different contributions \citep[e.g.,][]{delRio2004,Bouchard2010}. 
Similarly, the complex stellar and gas large-scale properties of NGC\,205 show signs of the past interactions, mainly with M\,31 \citep[e.g.,][]{Davidge2003a}, that would have led to the observed nuclear properties (see Appendix\,\ref{app:ind_gal} for more details and references). 
Although the nucleus of NGC\,5206 does not show such complex (kinematic) properties, previous studies have shown potential hints of past interactions \citep[e.g.,][]{Laurikainen2010}. These might have led to the formation of the NSC and its relatively low values of \vsRe\,and \lamRe.

\subsection{The formation of previously-studied NSCs}\label{sub:form_otherNSCs}
We add now to this discussion the NSCs from previous publications that were presented in Sect.\,\ref{sub:publ_NSCs}. 
\citet{Lyubenova2019} compared their kinematic results with $N$-body simulations from \citet{Antonini2012}, \citet{Perets2014}, and \citet{Tsatsi2017}. They showed that in general observed rotation can be recovered with multiple mergers of GCs, in a galactic center with an initially non-rotating bulge. 
Most of the results from \citet{Lyubenova2019} were in agreement with models of GCs infalling isotropically from random directions. These results  correspond to the open red circles in the bottom-left region of Fig.\,\ref{fig:vsig_comp} and \ref{fig:lamRe_comp}. 

However, in-situ star formation from infalling gas could also be the source of rotation in FCC\,47 and FCC\,170, with much higher values of \vsRe\,and \lamRe. 
While \citet{Lyubenova2019} showed that the high nuclear angular momentum of FCC\,47 could be explained with the infall of GCs from similar orbital directions to each other (as they would do if they formed in the galactic disk), 
a more complex origin was proposed by \citet{Fahrion2019}.
They invoked additional in-situ star formation, with mergers playing a significant role, to explain the high rotation, metallicity and mass ($\sim 7\times10^8$\msun) of this kinematically decoupled NSC. 
On the other hand, the position of FCC\,170 in the \lamRe\,diagram was not consistent with any of the GC-merger simulations including a bulge presented by \citet{Lyubenova2019}, who suggested that alternative mechanisms involving gas led to the formation and growth of this NSC. Therefore, in-situ formation would have been important in the four nuclei in early-type galaxies that are located in the upper region of Fig.\,\ref{fig:vsig_comp} and \ref{fig:lamRe_comp} (M\,32, NGC\,5102, FCC\,47 and FCC\,170).

The MW hosts the best studied NSC, whose origin is nevertheless still partly unclear due to its complex nature. 
On the one hand, it co-rotates with the Galactic disk but with a kinematic misalignment, while an additional inner component rotates perpendicular to the major axis. 
This structure suggests different episodes of star-cluster accretion \citep{Feldmeier2014,Feldmeier2017b}. 
On the other hand, the stellar populations disfavour a pure star-cluster infall scenario. 
For example, the wide range of metallicities is consistent with a formation in different episodes and probably via different mechanisms. 
The more metal-poor stars might have belonged originally to GCs that migrated to the Galactic center 
\citep{Feldmeier2017a,Feldmeier2020}. 

Furthermore, a kinematically-distinct metal-poor substructure has been recently identified in the central pc, probably as well a remnant of a massive star cluster or an accreted dwarf \citep{Do2020,ArcaSedda2020}.
However, the dominant populations, with their super-solar metallicity much higher than MW GCs, as well as a few very young stars uniformly concentrated in the center, were instead formed in situ 
\citep{Feldmeier2015}. 
A complex formation scenario, but with in-situ star formation contributing most of the mass and the strong rotation, might explain the location of the MW in the uppermost region of the \vsRe\,and \lamRe\,diagrams. 

The NSCs hosted by FCC\,47 and the MW, both massive (more than $10^7$\msun, see Sect.\,\ref{sub:publ_NSCs} for the MW), are represented by points close to the massive NSCs in our sample, in Fig.\,\ref{fig:vsig_comp} and \ref{fig:lamRe_comp}. Following the discussion in Sect.\,\ref{sub:form_fast} and in agreement with the mentioned previous studies, this supports a scenario in which a dominant role of gas inflow and in-situ star formation, contributing to higher values of \vsRe\,and \lamRe, is required for the growth of massive NSCs.

\subsection{Insights from globular clusters and ultra-compact dwarfs}\label{sub:gc_ucd}
The bottom-left region of Fig.\,\ref{fig:vsig_comp} and \ref{fig:lamRe_comp} is populated by nuclei of early-type galaxies, but also by the GCs of \citet{Kamann2018a}. 
These GCs do not extend to as high ellipticities, \vsRe\,or \lamRe\,values as the other NSCs, indicating that on average NSCs are more rotation dominated than the GC counterparts. 
On average, NSCs have larger masses than GCs, but low-mass NSCs have similar masses and metallicities to GCs at the high-mass end \citep{Fahrion2021}. 
If larger masses are related to stronger rotation (as suggested in Sect.\,\ref{sub:form_fast} and \ref{sub:form_otherNSCs}), massive GCs will have similar kinematic properties to less massive NSCs as we see in Fig.\,\ref{fig:vsig_comp} and \ref{fig:lamRe_comp}. 

\citet{Kamann2018a} suggested that GCs are born with significant angular momentum, inherited from the progenitor gas, which is however dissipated over time.
This is consistent with a recent study by \citet{Lahen2020}, on the kinematics of simulated young massive star clusters, formed during a merger-induced starburst. They showed how some recently formed massive clusters lie in the upper region of the \vsRe\,and \lamRe\,diagrams, where our late-type nuclei are located.
 \citet{Pfeffer2020} agreed that the observed properties of GCs can be the result of the evolution of (massive) star clusters, formed in the early Universe initially with similar properties as present-day young clusters (see also \citealt{Portegies2010}). 
Hence, not only NSCs can form from the infall of GCs to the galactic center, but also the seeds for the formation of both NSCs and GCs might have had very similar properties. 
 
In addition, as suggested e.g. by \citet{Boeker2008}, (some) GCs might be the nuclear remnants of dwarf satellites accreted by massive galaxies. 
This is another channel of the tight evolutionary connection between (some) NSCs and GCs.
Some MW GCs, in fact, are thought to be stripped nuclei. This is the case, e.g., of $\omega$\,Cen \citep{Hilker2000,Noyola2008,Pfeffer2021} and M\,54 \citep{Ibata2009,AlfaroCuello2019,AlfaroCuello2020,Pfeffer2021}, the two most massive GCs in the MW. 
They were not included in the sample by \citet{Kamann2018a}. 

M\,54, with available kinematics from \citet{AlfaroCuello2020}, was added in Fig.\,\ref{fig:vsig_comp} and \ref{fig:lamRe_comp}. 
M\,54 is the stripped nucleus of the disrupted Sagittarius dwarf spheroidal galaxy, as supported by its complex stellar populations \citep{Bellazzini2008,Mucciarelli2017,AlfaroCuello2019}. 
It lies as well in the bottom-left region of the diagrams. As shown by \citet{AlfaroCuello2019,AlfaroCuello2020}, its integrated kinematics is the result of the combination of three main stellar populations: an old and metal-poor component with a very low amount of rotation, an intermediate-age metal-rich component with some rotation, and a young, faster-rotating, metal-rich population. 
While properties of the young component pointed to an in-situ origin, GC accretion was proposed for the old populations. 
M\,54 is one more example, in the bottom region of the \vsRe\,or \lamRe\,diagrams, of a complex nuclear structure made up of components with different origin, probably the results of past interactions. 

We also include two stripped nuclei into our discussion, M\,59-UCD\,3 and M\,60-UCD\,1. Their central over-massive SMBHs suggest that these UCDs are the stripped nuclei of much more massive progenitors ($10^9-10^{10}$\msun) \citep{Seth2014,Ahn2018}.
The latter hosted NSCs, that are now observed as the inner morphological components of these UCDs (see also \citealt{Pfeffer2013}). 
Nevertheless, nothing can be said about the morphological type of these progenitors. 
These stripped nuclei are located, in Fig.\,\ref{fig:vsig_comp} and \ref{fig:lamRe_comp},  in the bottom-left region characterized by GCs and nuclei in early-type galaxies, very close to NGC\,205's nucleus.
Both show strong rotation and high velocity dispersion (with a central peak associated with the SMBH). 
M\,59-UCD\,3 displays a complex structure with multiple stellar populations \citep{Ahn2018}, similarly to other early-type galaxies in our sample (Sect.\,\ref{sub:form_hosts}). 
M\,60-UCD\,1 seems to be populated mainly by uniformly old stars and suggests that it was stripped a long time ago preventing any recent in-situ growth \citep{Seth2014}. 
However, the amount of rotation observed in the inner region of these peculiar objects ($\sim40$\kms, \citealt{Seth2014,Ahn2018}) is much larger than the one observed in most early-type nuclei in our sample. It is instead comparable to the ones located in the upper region of the \vsRe\,or \lamRe\,diagrams. 

\section{Conclusions}\label{sec:concl}
We used a unique data set, from AO-assisted IFS in the CO band head, of the nuclear regions of eleven early and late-type galaxies. 
From these data cubes, we extracted and analysed their resolved kinematics, at a parsec or subparsec scale. 
We provided the high-resolution maps of the first four moments of the LOSVD and their uncertainties, indicating that some level of rotation is ubiquitous in NSCs. 
The maps extracted from the highest-S/N data cubes, show some kinematic complexity. 

We analysed the balance of ordered rotation and random motions in our targets, making use of the \vsRe\,and \lamRe$ - \epsilon_{\rm{e}}$ diagrams. 
Nuclei hosted by late-type galaxies, 
are located in the upper, rotation-dominated region of the diagrams. 
On the other hand, early-type nuclei display a larger variety of cases corresponding to different locations in the diagrams. 
They show, on average, lower amount of rotation and lower ellipticity, very often associated to complex kinematic structures. 
However, some peculiar early-type galaxies host massive, strongly-rotating nuclei located in the region of the diagrams dominated by late types. 
Therefore, nuclei hosted by late-type galaxies and massive nuclei in early types populate the upper region of the \vsRe\,and \lamRe$ - \epsilon_{\rm{e}}$ diagrams. 

If rotation is associated with in-situ formation from gas accretion (or the infall of coplanar star clusters), and pressure support to the merging of star clusters (in general from random directions), the continuous distribution of our points in the \vsRe\,and \lamRe$ - \epsilon_{\rm{e}}$ diagrams supports complex formation pictures driven by a mix of these scenarios. Since the highest values of \vsRe\,and \lamRe\,correspond to nuclei in late-type galaxies and massive nuclei in early types, we suggest that in-situ formation is the dominant NSC formation mechanism in late-type galaxies, but also required for the growth of massive NSCs in galaxies of all types.
This picture is supported by additional samples from previous studies.

The discussion of the properties of the individual nuclei in the context of their hosts supports our conclusions and suggests that in general galaxy interactions, often triggering gas inflow and star-formation bursts, but also contributing complex structures, play a fundamental role in the NSC formation. 
The variety of nuclear kinematic properties in early-type galaxies is probably closely connected to the specific evolution history of their hosts, which can
therefore form their NSCs in quite different ways. 

Our work provides a first glimpse into the internal kinematics of a sample of NSCs,  suggests that rotation is ubiquitous and provides information on the balance between the two most invoked formation scenarios proposed for NSCs. 
However, the variety of specific cases, complex structures and peculiar galaxy-evolution histories make it challenging to reconstruct the NSC formation only by kinematic results. 
Spatially resolved stellar-population properties, including star-formation histories, would be necessary to understand better the complexity of these NSCs. 
Moreover, 
connecting the kinematics to detailed information on the formation mechanism is limited by seeing, projection effects and the contributions of the underlying galaxies. Detailed modeling of these data including these effects more fully has the potential to yield better understanding of NSC formation mechanisms. 

\acknowledgments
We would like to thank Anais Gonneau for her advise on the use of the XSL and Jesús Falcón-Barroso for the useful technical discussions. 
A special thank to Anja Feldmeier, Mariya Lyubenova and Mayte Alfaro-Cuello for kindly providing the tables of their published results cited here. 
We also thank the anonymous referee for their valuable comments.

\facilities{Gemini:Gillett (NIFS/ALTAIR), VLT:Yepun (SINFONI/AO)}

\software{Gemini IRAF Package v 1.9 (\url{https://www.gemini.edu/observing/phase-iii/understanding-and-processing-data/data-processing-software/gemini-iraf-general}); 
IRAF \citep{Tody1986, Tody1993};
PampelMuse \citep{Kamann2018b}; 
ESO SINFONI pipeline \citep{Eisenhauer2003, Freudling2013};
VorBin \citep{Cappellari2003};
pPXF \citep{Cappellari2004, Cappellari2017}; 
MgeFit \citep{Emsellem1994, Cappellari2002};
astropy \citep{astropy2013, astropy2018};
photutils \citep{Bradley2020}.
}

\newpage
\bibliography{fp_nsc}{}
\bibliographystyle{aasjournal}

\newpage
\appendix

\section{Individual nuclei: their host galaxies and kinematic maps} \label{app:ind_gal}
In this Appendix, we describe briefly the host galaxies in our sample, giving the potentially relevant information for the formation of their NSCs, and 
we show individual nuclear-kinematic maps extracted from our observations. 
In the top row of Fig.\,\ref{fig:m32} to \ref{fig:ngc7793} we show the first four moments of the LOSVD. From left to right: mean velocity $V$, velocity dispersion $\sigma$, skewness $h_3$ and kurtosis $h_4$. 
$V$ (top-left panel) was corrected for an approximation of the systemic velocity, calculated as the average value within the central 0.3\,arcsec\,$\times$\,0.3\,arcsec. 
In the bottom row, we show the respective uncertainties, per Voronoi bin, of the first four moments of the LOSVD. 
Dashed green ellipses indicate the \reff\,elliptical isophotes.  
For completeness and to avoid a subjective selection, we show the full Voronoi-binned area, although numerous bins display large uncertainties. These give often a qualitative idea of the kinematic trend in the outer region of the nuclei. 
Similar kinematic maps of M\,32, NGC\,205, NGC\,404, NGC\,5102 and NGC\,5206 were already published \citep{Seth2008b,Seth2010b,Seth2010a,Nguyen2018}. 
However, $h_3$ and $h_4$ were not included for all of them and, when possible, we cover a wider area. 
For these reasons and for completeness, we show here the maps of our full sample, consistent 
with the previously-published ones. 

\subsection{Early-type galaxies}\label{sub:early}
Of the eleven host galaxies in our sample, five are classified as early types (Table\,\ref{tab:sample}). We describe and discuss briefly each one of them and the kinematics of their nuclei as follows.

\subsubsection{M\,32} \label{subsub:m32}
M\,32 is a dwarf compact elliptical (cE) in the Local Group, close satellite of M\,31. 
It hosts the least massive of the three supermassive black holes (SMBHs) detected in the Local group ($ \sim 2.4 \times 10^6$\msun) 
\citep[e.g.,][]{Verolme2002,vandenBosch2010,Seth2010a,Nguyen2018}.
M\,32's origin is still controversial. As also suggested by its SMBH, it is thought to be a tidally-stripped remnant, result of the interaction with its massive neighbour M\,31 \citep[e.g.,][]{Dierickx2014,dSouza2018}. 
One possibility is that its progenitor was a low-luminosity spiral \citep{Bekki2001}. 
In support of this \enquote{threshed-spiral} scenario, a faint stellar disk was identified by \citet{GrahamA2002} and \citet{Nguyen2018} in the outer parts of M\,32. 
Furthermore, stellar kinematics of M\,32 from \citet{Verolme2002} and \citet{Dressler1988} 
show disk-like rotation with a maximum velocity of $\sim 60$\kms, slightly decreasing towards the outskirts. 
Two main stellar populations were identified in M\,32. One older than 5\,Gyr with subsolar metallicities, the other younger and metal rich \citep{Schiavon2004,Monachesi2012}, contributing mostly in the nuclear region 
\citep{Rose2005,Coelho2009,Miner2011}.

\begin{sidewaysfigure}[ht]
\centering
\resizebox{0.9\textwidth}{!}
{\includegraphics[scale=1, trim={1.3cm 0 0 0}]{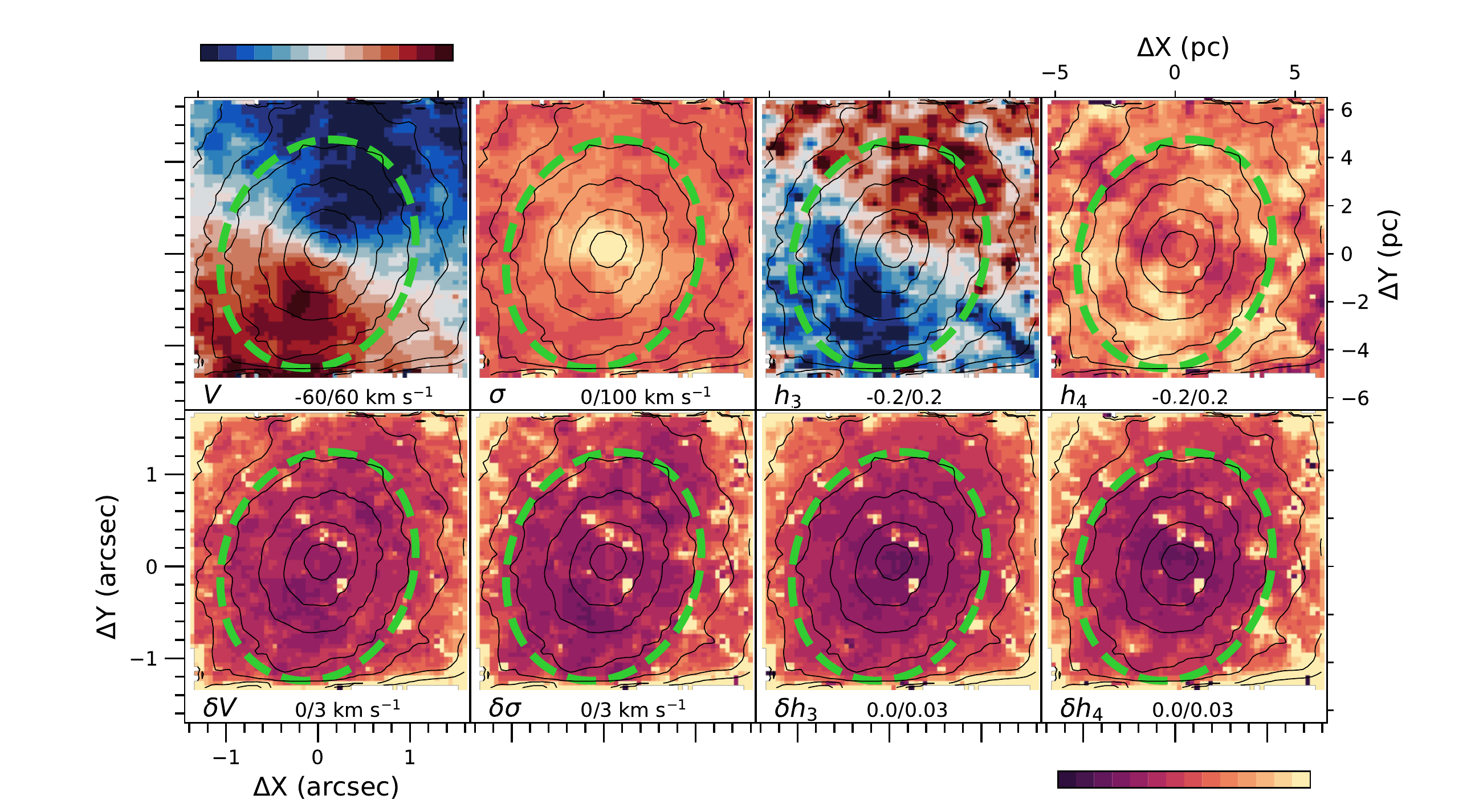}}
\caption{For the nuclear region of M\,32, maps of the first four moments of the LOSVD, in the top row. From left to right, as indicated on bottom left of each panel: mean velocity $V$, velocity dispersion $\sigma$, skewness $h_3$ and kurtosis $h_4$. Their respective uncertainties are shown in the bottom row. The color bar shown on top left of the figure was used for $V$ and $h_3$, while the one indicated on bottom right of the figure was used for $\sigma$, $h_4$ and all uncertanties. 
The maximum and minimum values of the color bars for each map are indicated at the bottom of each panel. 
The original data cube was Voronoi-binned to a signal-to-noise ratio (S/N) of 25 per bin.
Different isophotes are plotted in black to guide the eye. 
The dashed green ellipse corresponds to the \reff\,elliptical isophote. 
}
\label{fig:m32}
\end{sidewaysfigure}

The high S/N per spaxel allows us to observe kinematic features in M\,32's nucleus at sub-parsec resolution in the full FoV, as shown in Fig.\,\ref{fig:m32}. 
We find evidence of a relatively fast-rotating disk-like component, with velocities up to $\sim 60$\kms\,(top-left panel) that are clearly anticorrelated with skewness (third top panel from the left). This anticorrelation suggests disk-like rotation superimposed on weaker or no rotation \citep[e.g.,][]{vanderMarel1993}. 
Velocity dispersion displays a strong peak in the center, which can be explained by the presence of a SMBH. 
A structure of higher values of $\sigma$ appears to be aligned with the minor axis.
The kurtosis map, anticorrelated with $\sigma$, shows slightly negative values, indicating broader profiles of the LOSVD with respect to a Gaussian, along the rotation axis and positive values across the rest of the disk. 
As discussed by \citet{Seth2010a}, these $h_3$ and $h_4$ maps suggest the superposition in the LOS of a dominant rotating disk on a much slower or non-rotating bulge.

Due to the proximity of this galaxy, small structures in the eight maps might correspond to individual stars, 
deviating from the integrated kinematics of the galaxy and determining larger uncertainties. 
The high quality of this data allows us to determine the kinematic parameters with low uncertainties: generally below 2\kms\,for $V$ and $\sigma$ and below 0.02 for $h_3$ and $h_4$. Uncertainties are higher close to the edges of the FoV.
Kinematic maps in \citet{Seth2010a} are within our uncertainties. Our results are also compatible with other previous studies \citep{Dressler1988,Joseph2001}. 

The strong disk-like rotation in the massive M\,32's nucleus ($\sim 10^7$\msun, see \citealt{Nguyen2018}), typical of 
the central regions of disky elliptical galaxies, is consistent with an in-situ formation from accreted gas 
\citep{Joseph2001,Hoffman2009,Seth2010a}. 
Nuclear kinematics can be related to M\,32's controversial origin. 
In Fig.\,\ref{fig:vsig_sample} and \ref{fig:lamRe_sample}, the nucleus of M\,32 lies in the uppermost region covered by late-type galaxies and shows one of the highest proportion between ordered and random motions, in spite of the high values of velocity dispersion. 
If the compact galaxy that we see today is a tidally stripped remnant, the nucleus might be still bearing the footprint of its \enquote{threshed}-spiral host. It corrotates with the galaxy, as from comparing with the larger-scale kinematics mapped by \citet{Verolme2002}, while its chemical properties, consistent with a spiral and not with a classical elliptical galaxy, point as well to this same scenario \citep{Davidge2008a}. 
Strong tidal interactions with M\,31 may have 
also 
favoured gas inflow towards the center, ending in a starburst forming the youngest populations in the NSC \citep{Bekki2001}.

\subsubsection{NGC\,205}\label{subsub:m110}
NGC\,205 is also located in the Local Group and is the brightest and the closest to M\,31 of its three dwarf-elliptical (dE) companions \citep{Davidge2005}. Several signs of past interactions with M\,31 have been found in both stellar and gas properties of NGC\,205 \citep[e.g.,][]{Cepa1988,Young1997,Mateo1998,Davidge2003a,McConnachie2004,Thilker2004}.
NGC\,205 was initially thought to be globally an old galaxy, since its stellar populations were associated with the ones in MW GCs   \citep{Baade1944,Mould1984}.
However, signs of recent star-formation activity, probably triggered by interactions with M\,31, were later found in the central regions \citep{Mateo1998,Cappellari1999, Davidge2003a,Butler2005}.

\begin{sidewaysfigure}[ht]
\centering
\resizebox{1.01\textwidth}{!}
{\includegraphics[scale=1.6, trim={1.3cm 0 0 0}]{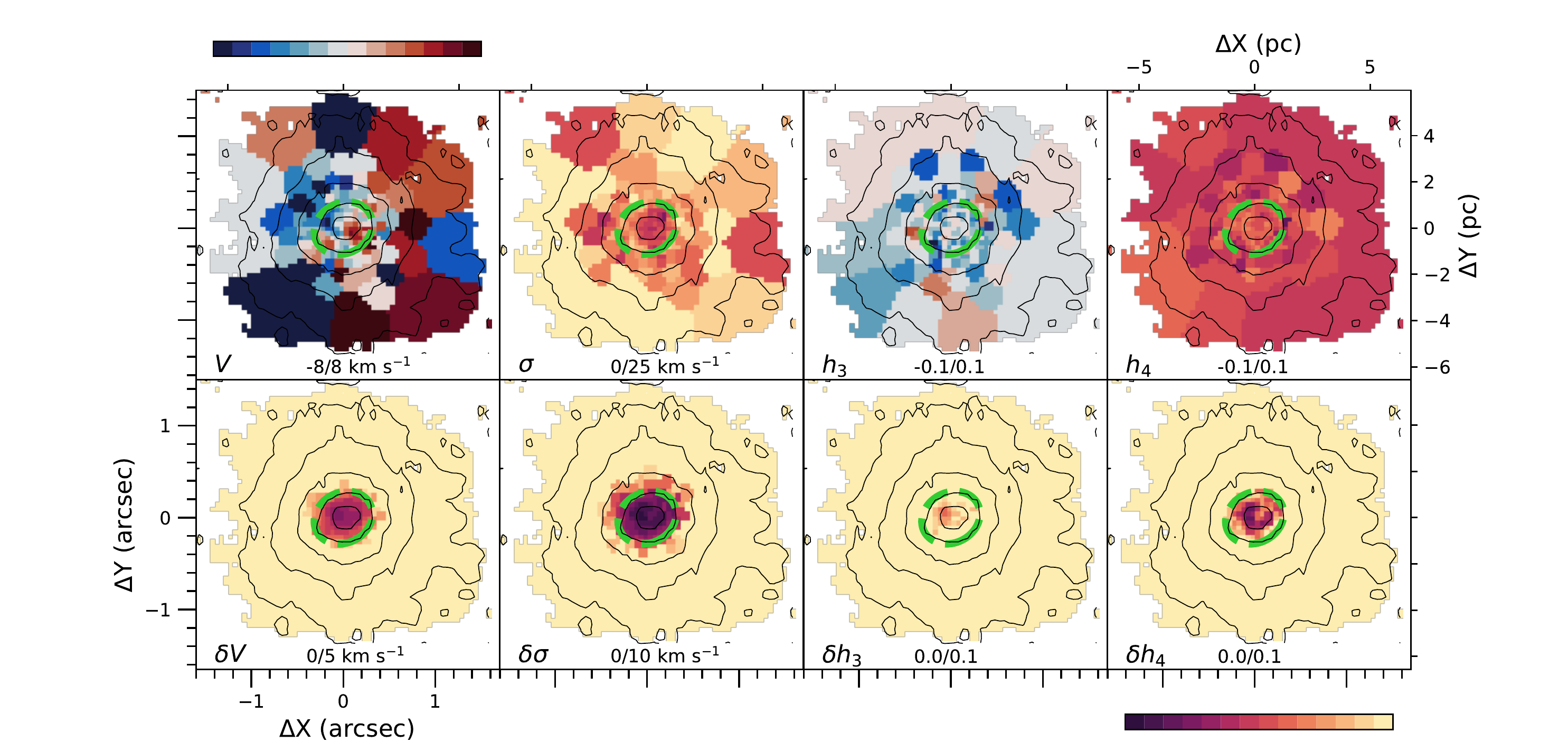}}
\caption{As Fig.\,\ref{fig:m32}, now for the nuclear region of NGC\,205. 
The original data cube was Voronoi-binned to a signal-to-noise ratio (S/N) of 25 per bin.
}
\label{fig:ngc205}
\end{sidewaysfigure}

We observe relatively slow rotation in the nucleus of NGC\,205. 
The rotation does not happen around the photometric minor axis, as seen in the top-left panel of Fig.\,\ref{fig:ngc205}, suggesting that mergers played an important role in the formation of this NSC. 
Velocity dispersion drops in the central region within the \reff\,elliptical isophote,
in agreement with the presence of a low-mass central BH as first investigated by \citet{Valluri2005} and \citet{Nguyen2018}.  \citet{Nguyen2019} provided a model of the measured kinematics and measured the BH mass 
($\gtrsim 5 \times 10^3$\msun). The low $\sigma$ values are also consistent with the observed dominant young dynamically-cold populations \citep{deRijcke2006}. 

We consider reliable the absolute values of $V$ and 
$\sigma$ only within the $0.5 - 0.7$\,arcsec isophotes, where uncertainties are not too large. However, a larger portion of the FoV hints that the galaxy may follow a similar rotation pattern and a velocity dispersion increasing farther from the center.
Significant rotation was measured at larger scales, but along the major axis of the galaxy \citep{Simien2002}.
No pattern can be distinguished in our $h_3$ and $h_4$ maps, in general with large uncertainties. 
Our results are consistent with the kinematic maps published by \citet{Nguyen2018}. 

The nucleus of NGC\,205 is located, in Fig.\,\ref{fig:vsig_sample} and \ref{fig:lamRe_sample}, in the bottom-left region dominated by nuclei in early-type galaxies. 
Although this might be partially due to the impact of the PSF (see Appendix\,\ref{app:PSF}), 
we suggest that it is also related to the complex structure suggested by both the kinematics and the stellar populations that were found in this blue NSC, whose total mass is $\sim 2\times 10^6$\msun\, \citep{deRijcke2006}. 
Young stellar populations ($<1$\,Gyr old) dominate, while they coexist with intermediate-age stars \citep{Bica1990, Butler2005,Monaco2009,Nguyen2019}, suggesting that this NSC was formed in different episodes and different mechanisms might have been at play. Its rotation decoupled from its larger scale host, points to a significant merger contribution for its stars or the gas originating them.
\begin{sidewaysfigure}[ht]
\centering
\resizebox{1.01\textwidth}{!}
{\includegraphics[scale=1.6, trim={1.3cm 0 0 0}]{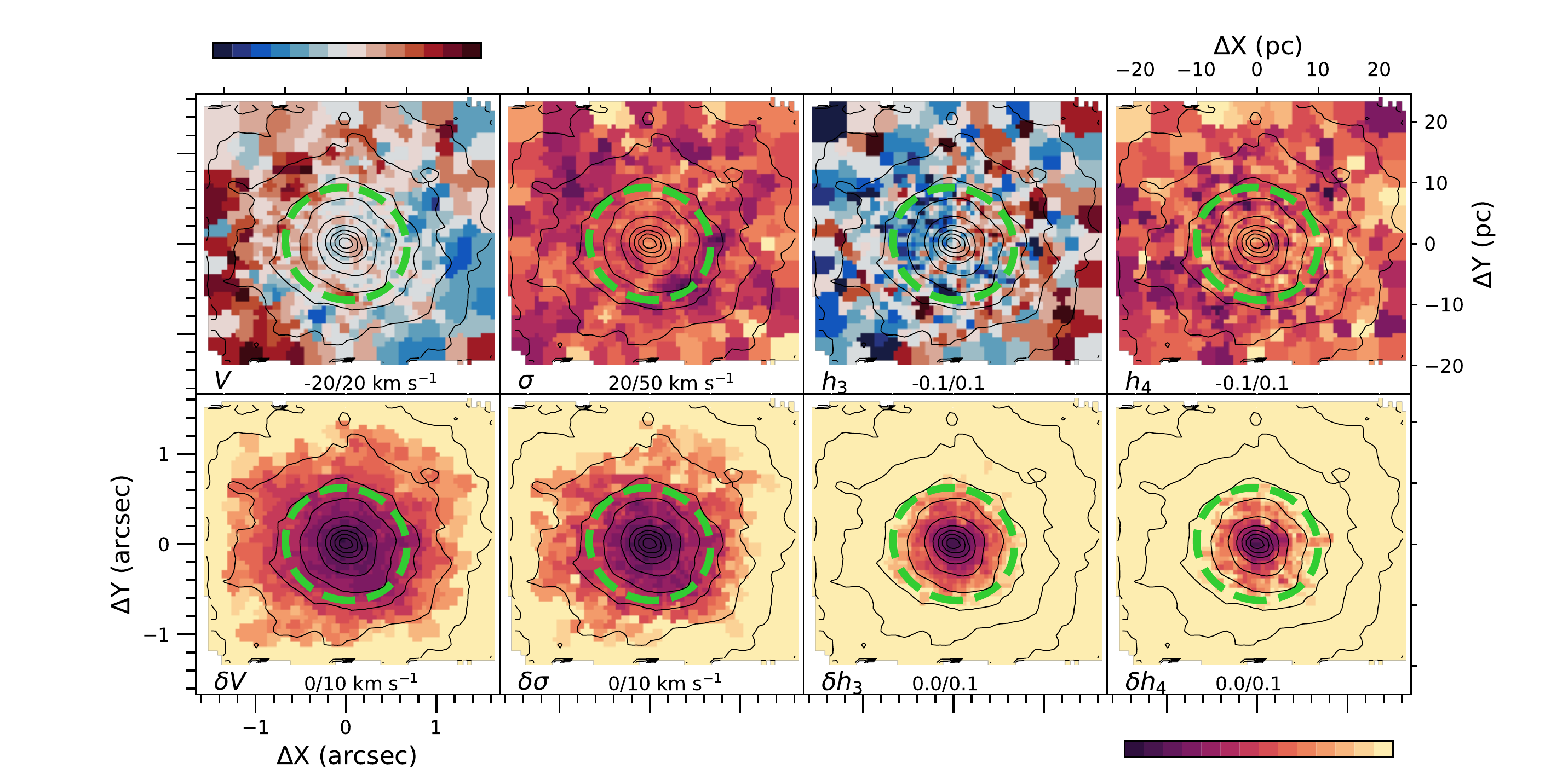}}
\caption{As Fig.\,\ref{fig:m32}, now for the nuclear region of  NGC\,404. 
The original data cube was Voronoi-binned to a signal-to-noise ratio (S/N) of 25 per bin.
}
\label{fig:ngc404}
\end{sidewaysfigure}
\subsubsection{NGC\,404}\label{subsub:ngc404}
NGC\,404 is a nearly face-on dwarf lenticular galaxy ($\sim11\deg$ inclined, \citealt{delRio2004}), 
hosting a low-ionization nuclear emission-line region (LINER)  \citep{Ho1997,Dumont2020,Boehle2018}, powered by a BH of $\sim 5.5\times10^5$\msun \citep{Davis2020}. Member of the galaxy group LGG\,11, it is relatively isolated \citep{Garcia1993,Williams2010b}.
NGC\,404 is characterized by a prominent warped HI component, something rather exceptional for an early-type galaxy. 
An extended gaseous disk with a doughnut shape, with the optical galaxy located within the hole, is combined with a misaligned and kinematically-decoupled more elliptical annulus \citep{delRio2004}.

A merger origin was proposed for this complex structure, with recent star formation associated with a rejuvenating process (see also \citealt{Bouchard2010,Williams2010b,Thilker2010}).
The properties in the central region of this galaxy, with a massive gas and dust cloud and younger stellar populations than in the disk, 
support this scenario and suggests a star formation in different episodes \citep{Wiklind1990,Tikhonov2003,CidFernandez2005,Bouchard2010,Williams2010b, Seth2010b}.
The complex stellar structure of NGC\,404 may be the result of the combination of in-situ older components with newer ones acquired during mergers. These may have taken the galaxy through a morphological transition from spiral to lenticular \citep{Bouchard2010}. 

The kinematics of NGC\,404's nucleus is shown in Fig.\,\ref{fig:ngc404}. 
We obtained good $V$ and $\sigma$ maps in a large portion of the FoV, recovering the same structures and values (within our errors) as published by \citet{Seth2010b}. 
We add here $h_3$ and $h_4$, with low uncertainties only within the \reff\,elliptical isophote. 
Kinematics reveals a complex structure. Out of the \reff\,ellipse, we observe a disk-like rotating structure, as also suggested by the $V-h_3$ anticorrelation. This component is kinematically decoupled from the close surroundings, but co-rotating with the HI disk \citep{delRio2004,Bouchard2010, Seth2010b}. 
The region within the \reff\,rotates slower and/or the lower measured $V$ might be due to the additional counter-rotating structure that is seen in the center of our $V$ map. 
This was already identified by \citet{Seth2010b} within the central $\sim 0.2$\,arcsec and associated with a distinct photometric component and the detection of hot dust. 
A higher $\sigma$ within the \reff\,than in the surroundings is observed. The central values are in agreement with \citet{Ho2019}. 

The presence of the inner counter-rotating component superposed to the outer one leads to a lower integrated rotation velocity and a higher dispersion in the LOS. 
In fact, NGC\,404 is the lowest and left-most point in the diagrams in Fig.\,\ref{fig:vsig_sample} and \ref{fig:lamRe_sample}. 
While its location is probably affected by different factors such as the low inclination, the contamination of the underlying galaxy and 
the mass of the central BH, it might be mainly due to the presence of kinematically distinct components in the LOS (see also Sect.\,\ref{sub:vslam_disc}). 

A dominant stellar age of $\sim 1$\,Gyr and some other younger and older populations, suggest the  formation of this NSC during bursts of star formation triggered by gas inflow during mergers \citep{CidFernandez2005,Seth2010b,Bouchard2010}.
\citet{Nguyen2017} suggested that the 1\,Gyr population is most dominant in the central $\sim 0.2$\,arcsec, indicating that the counter-rotating component was formed at that time, probably in situ after a minor merger, as suggested by its compactness. 
NIR emission lines in this nucleus revealed also gas thermal excitation from shocks \citep{Seth2010b,Boehle2018}. 
The complex gas structure of the galaxy also points to the merger scenario. 
In conclusion, the position of this nucleus in the \vsRe\,and \lamRe\,diagrams might be indicating a complex NSC formation, led by 
past mergers either via star-cluster accretion or in-situ starbursts after gas inflow.

\subsubsection{NGC\,5102}\label{subsub:ngc5102}
NGC\,5102 is located in the Cen\,A group and it is classified as an S0, although some of its properties are not typical of an early type \citep[e.g.][]{Davidge2008b}. 
For instance, it is bluer than a typical S0 and it has an extended H\,I disk \citep{vanWoerden1993}.
\citet{Davidge2008b} defined it as a post-starburst galaxy because of the signs of past large-scale star formation, and he suggested that it could have been a late-type spiral in the past.
Integral-field spectroscopy data from MUSE \citep{Mitzkus2017}, covering up to the galaxy effective radius, revealed two counter-rotating disks, one more centrally concentrated than the other. 
\citet{Mitzkus2017} argued that the more extended disk had been formed already, when counter-rotating gas was accreted and formed the second disk (see also \citealt{vanWoerden1993}). 
Stellar-population analysis shows a strong gradient in both age and metallicity, with younger and more metal-rich stars towards the center. 
X-ray emission was detected by \citet{Kraft2005} both as a point source in the center probably indicating a low-luminosity active galactic nucleus (LLAGN), and a diffuse emission, in the central kpc, probably due to hot gas shocked during the most recent starburst.
A SMBH (M$_{\rm{BH}} \sim 9\times 10^5$\msun) was detected by \citeauthor{Nguyen2018} (\citeyear{Nguyen2018,Nguyen2019}). 

\begin{sidewaysfigure}[ht]
\centering
\resizebox{0.9\textwidth}{!}
{\includegraphics[scale=1.6, trim={1.3cm 0 0 0}]{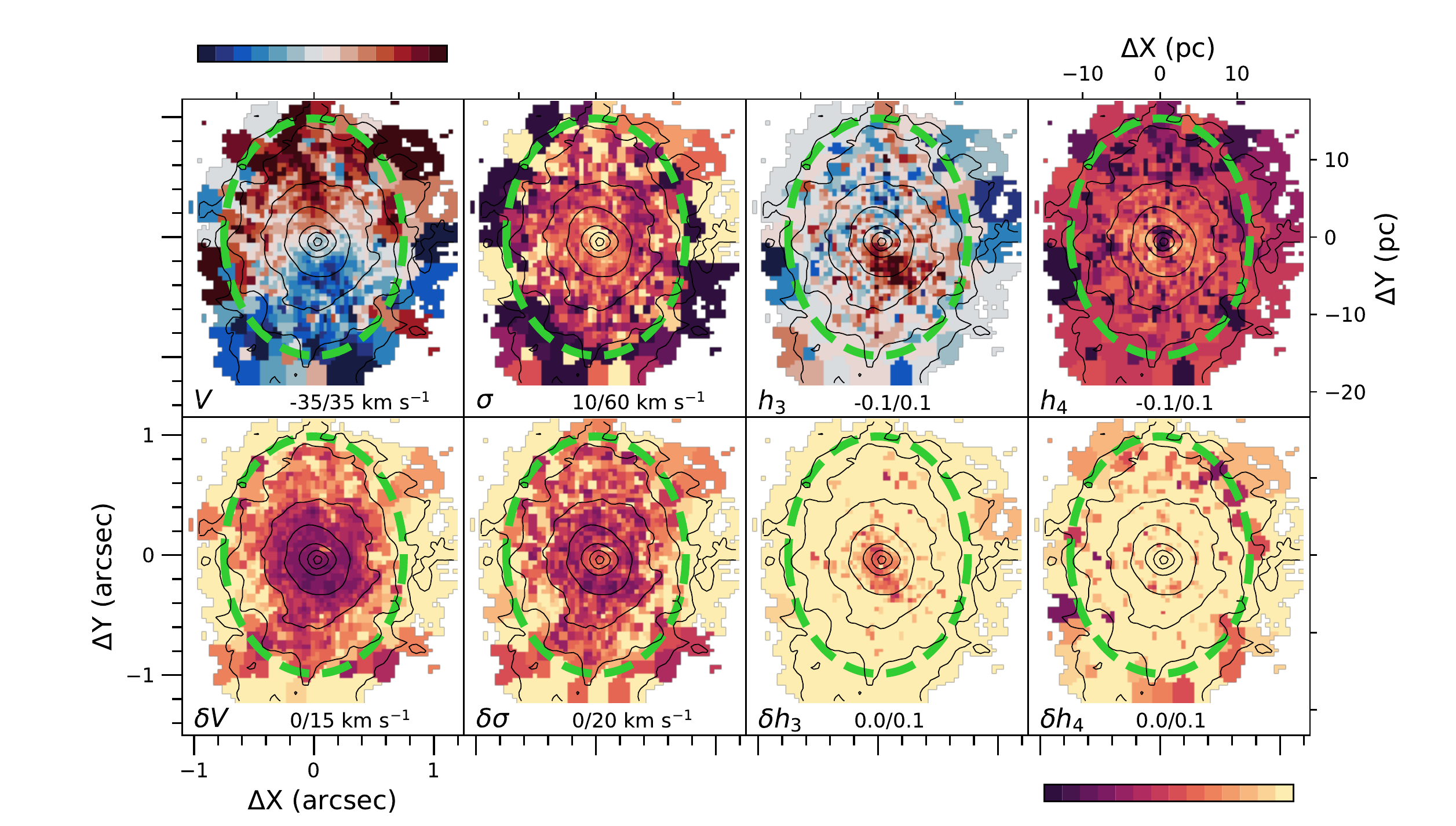}}
\caption{As Fig.\,\ref{fig:m32}, now for the nuclear region of NGC\,5102. 
The original data cube was Voronoi-binned to a signal-to-noise ratio (S/N) of 25 per bin.
}
\label{fig:ngc5102}
\end{sidewaysfigure}

This NSC was photometrically fitted by \citet{Nguyen2018} with two S\'ersic components. 
One of them is more massive, flatter and more extended (with \reff$\sim 32$ pc) than the inner younger one (with \reff$\sim 1.6$ pc). 
Not the entire outer component was covered by the SINFONI FoV. 
In Fig.\,\ref{fig:ngc5102} we indicate with a green dashed ellipse the area where we integrated \vsRe\,and \lamRe\,(see Sect.\,\ref{subsec:vsigma}).
The region in our FoV displays relatively strong rotation, with maximum values at almost 40\kms\,(top-left panel). 
This rotation corresponds to the outer S\'ersic component. 
\citet{Nguyen2018} suggested that both components may have been formed in situ, in different episodes, from gas inflow related to mergers.
\citet{Davidge2015}, \citet{Mitzkus2017} and \citet{Kacharov2018} measured ages of $\lesssim 1$\,Gyr in the NSC, much younger and more metal poor than the surrounding galaxy. 
The nuclear outer component co-rotates with the larger-scale inner disk found by 
\citet{Mitzkus2017}, and both counter-rotate with respect to the outer galaxy. Therefore, both may have formed from the same gas-accretion processes. 

The very central region is associated with a peak in velocity dispersion up to $\sim 60$\kms\,(second top panel from left), consistent with the presence of a SMBH.
$h_3$ and $h_4$ are anticorrelated respectively with $V$ and $\sigma$, although their values show large relative uncertainties.
This NSC appears to be more massive than the average for galaxies with similar stellar mass or velocity dispersion, with a NSC stellar mass of $\sim 7 \times 10^7$\msun \citep{Nguyen2018}. Like the massive one in M\,32, it lies in the upper region of the \vsRe\,and \lamRe\,diagrams (Fig.\,\ref{fig:vsig_sample} and \ref{fig:lamRe_sample}), more typical of late-type hosts. 
Both galaxies may show, in their nuclear kinematics, hints of their past as spirals and of the interactions that led to their transformation to early types.

\subsubsection{NGC\,5206}\label{subsub:ngc5206}
NGC\,5206 is another nucleated S0 galaxy in the Cen\,A group. 
Not much is found in the literature about the past of this galaxy. However, the following complex structure may be the signature of past interactions. 
Apart from the bright nucleus \citep{Caldwell1987}, its brightness profile may indicate the presence of a bulge-like component, with a bulge-to-total flux ratio of 0.08, closer to values of typical spiral galaxies \citep{Laurikainen2010}.
Moreover, \citet{Laurikainen2010} described a \enquote{very faint dispersed bar} (at radii lower than 1.4\,kpc) and a faint lens-like structure at radii lower than 0.3\,kpc. 
In contrast, \citet{Nguyen2018} fitted HST images with only one (disk) component apart from the NSC. They observed a color gradient towards bluer stars in the center. 
Some ionized gas was detected in this galaxy \citep{Cote1997,Kennicut2008}, and a poor atomic gas fraction (lower than 0.1\%) was calculated by \citet{deSwardt2010}. 
\citeauthor{Nguyen2018} (\citeyear{Nguyen2018,Nguyen2019}) detected a SMBH of M$_{\rm{BH}} \sim 5-6\times 10^5$\msun. 

\begin{sidewaysfigure}[ht]
\centering
\resizebox{1.01\textwidth}{!}
{\includegraphics[scale=1.6, trim={1.3cm 0 0 0}]{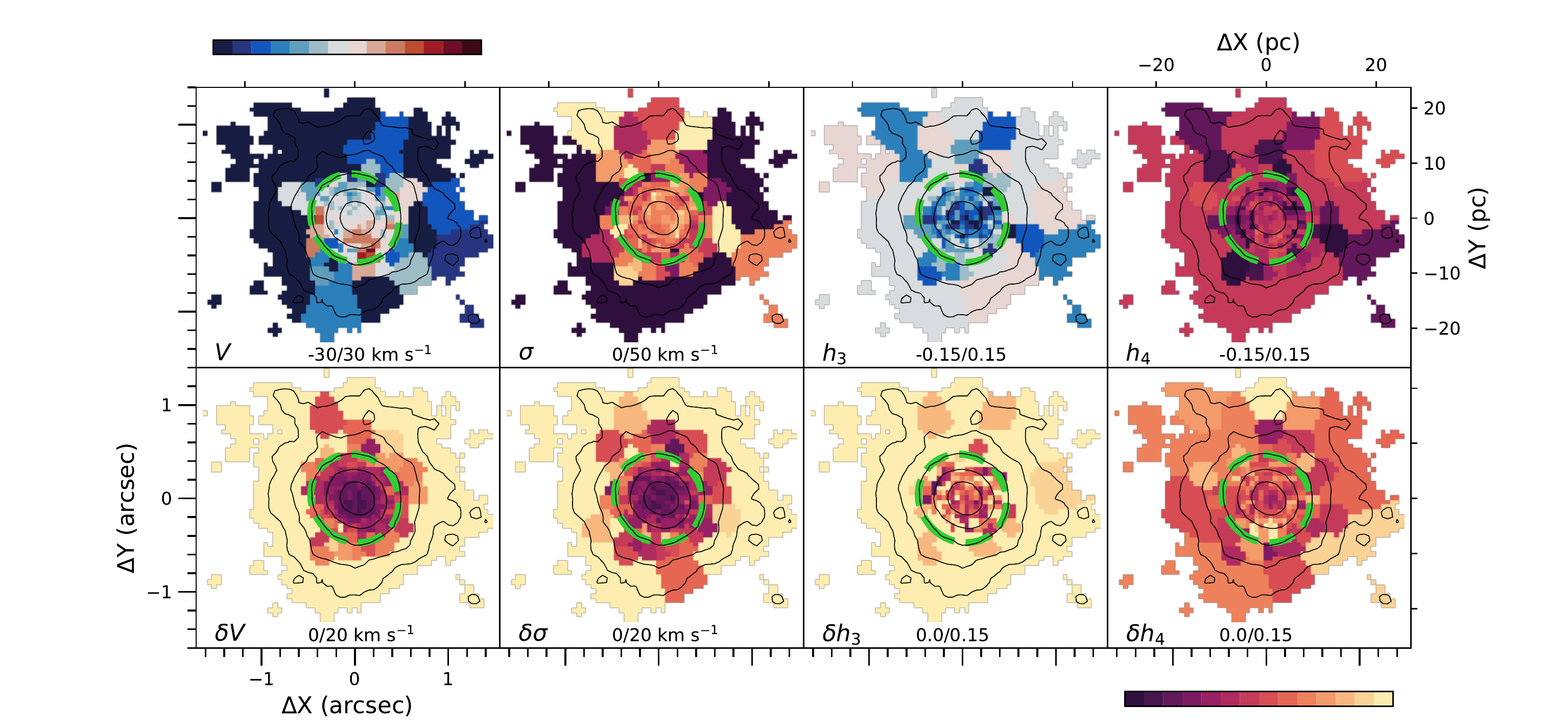}}
\caption{As Fig.\,\ref{fig:m32}, now for the nuclear region of NGC\,5206. 
The original data cube was Voronoi-binned to a signal-to-noise ratio (S/N) of 25 per bin.
}
\label{fig:ngc5206}
\end{sidewaysfigure}

Nuclear kinematic maps of NGC\,5206 (Fig.\,\ref{fig:ngc5206}) are characterized by slow rotation and higher velocity dispersions in the very central region, closer to the SMBH. 
\citet{Nguyen2018} presented kinematic maps in a smaller central region, with results consistent with ours. 
No structures can be identified in our maps of $h_3$ and $h_4$.
As in NGC\,5102, the brightness profile of the NSC was fitted with a double S\'ersic \citep{Nguyen2018}. 
However, in this case, the two components have similar morphologic and kinematic properties.
\citet{Kacharov2018} analysed the stellar populations of this NSC and proposed a continuous star formation and gradual chemical enrichment. 
Most NSC stars would have been formed less than 4.5\,Gyr ago, 
with a peak $\sim 2$\,Gyr ago. No substantial difference in age and metallicity was found with respect to the surrounding field stars of the galaxy.
As NGC\,205, it is located, in Fig.\,\ref{fig:vsig_sample} and \ref{fig:lamRe_sample}, in the bottom-left region but close to the \enquote{late-type dominated region}. 
In both galaxies, this might be related to the composition of different stellar populations with different kinematic signatures. 

\newpage
\subsection{Late-type galaxies}\label{sub:late}
Six host galaxies of our eleven nuclei are classified as late types (Table\,\ref{tab:sample}). Host galaxies and nuclear kinematics are individually described and briefly discussed as follows.

\subsubsection{M\,33}\label{subsub:m33}
M\,33 is a late-type spiral in the Local Group, 
thought to be orbiting around M\,31 \citep[e.g.,][]{vanderMarel2008}. 
Weak tidal interactions with M\,31 were proposed to explain different atomic and ionized-gas structures, such as extended disk warps, an arc, and a filament 
\citep[e.g.,][]{Corbelli1989,Corbelli1997,Putman2009, Corbelli2014, Semczuk2018,Tachihara2018}. 
The gas in these features, coming from a galactic fountain and/or being previously stripped from M\,33's disk in a previous interaction, is now falling back, fuelling the intense ongoing star formation \citep{Putman2009,Zheng2017}.

With no observed bulge, a halo component 
and a kinematically-distinct stellar stream were detected by \citet{McConnachie2006}. 
The galaxy disk shows stellar-population gradients that are consistent with an inside-out formation scenario \citep{Beasley2015, Mostoghiu2018}. However, these gradients are inverted 
outside the star-forming disk, suggesting a different (outside-in or ex-situ) origin (see also \citealt{Davidge2003b} and \citealt{RoblesValdez2013}). 
No signatures were found of any SMBH in M\,33 \citep[e.g.,][]{Gebhardt2001}.
However, it hosts a bright X-ray and radio source in its center 
\citep{Long1981,White2019} and 
it has been debated whether this emission is associated with a central (low-mass, $\lesssim 3000$\msun) BH \citep[e.g.,][]{Gebhardt2001,Merritt2001}. 
\begin{sidewaysfigure}[ht]
\centering
\resizebox{1.01\textwidth}{!}
{\includegraphics[scale=1.6, trim={1.3cm 0 0 0}]{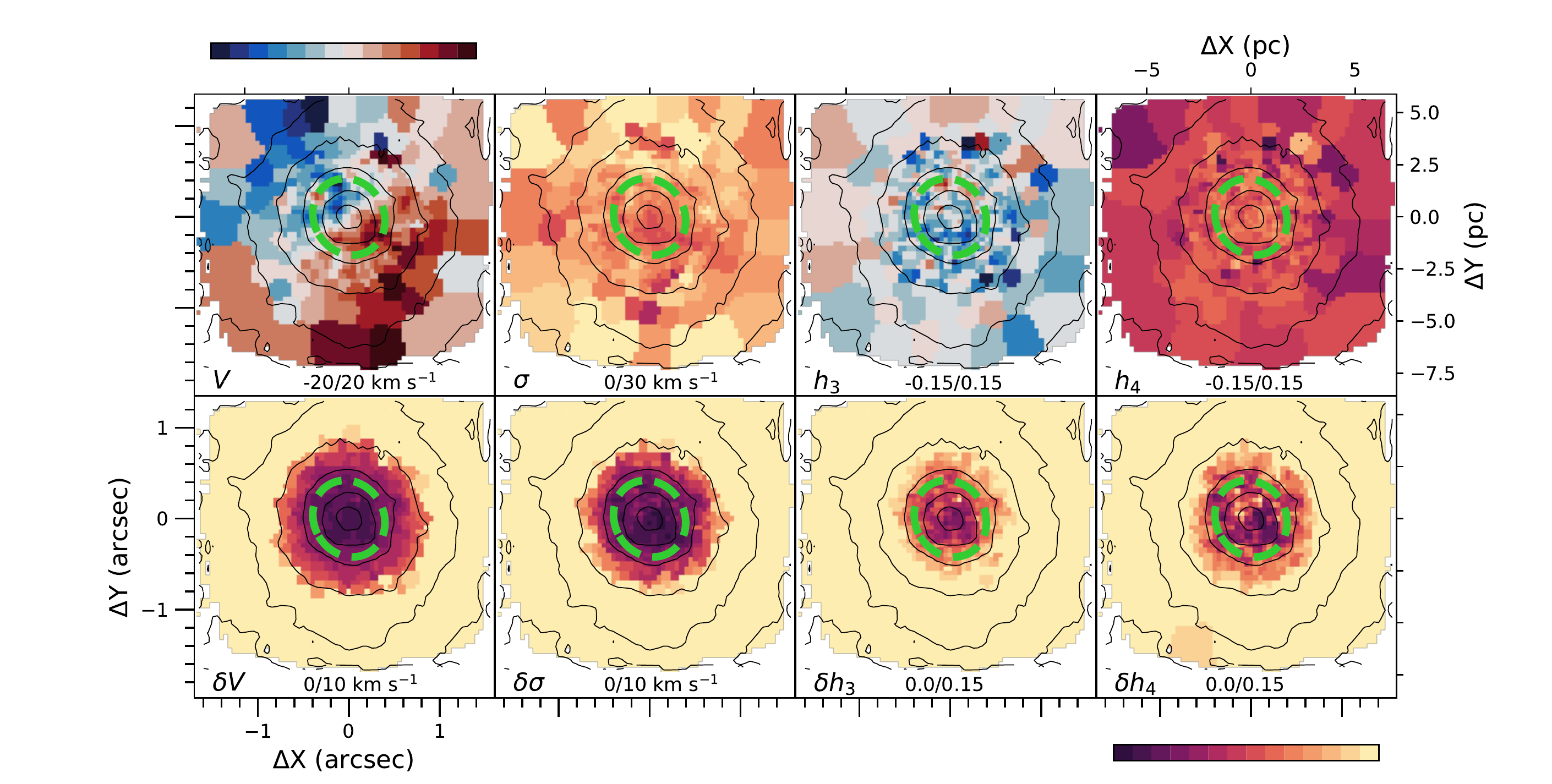}}
\caption{As Fig.\,\ref{fig:m32}, now for the nuclear region of M\,33.
The original data cube was Voronoi-binned to a signal-to-noise ratio (S/N) of 25 per bin.
}
\label{fig:m33}
\end{sidewaysfigure}

Maps of the kinematics of the nucleus in M\,33 are shown in Fig.\,\ref{fig:m33}. 
Very high spatial resolution is maintained in the (almost unbinned) central region of M\,33's nucleus, where uncertainties are below 5\,\kms.
Clear rotation is shown throughout the FoV, while the velocity dispersion drops within \reff, with values consistent with \citet{Kormendy1993} and \citet{Gebhardt2001}. 
This kinematics supports the lack of a central SMBH and is compatible with a low-mass central BH.
A $V-h_3$ anticorrelation is hinted while no structures are visible in the $h_4$ map.
The subtraction of individual stars from M\,33's data cube might be not perfect and leads to some residual structures that are seen in the maps.

The nucleus of M\,33, with a dynamical mass of $\sim 10^6$\msun\,\citep{Kormendy2010}, was defined as very compact.  It is characterized by younger stars than the surroundings and bluer colors towards its center \citep{Kormendy1993,Lauer1998,Carson2015}. 
It shows an important amount of dust, compatible with one or more strong starbursts occurred in the last Gyr \citep{Gordon1999,Davidge2000,Long2002}, probably fuelled by the observed gas infall.
\citet{Hartmann2011}, simulating the accretion of young stellar clusters to an in-place nuclear disk,  could recover the properties of the NSC in M\,33 (rotation, size, ellipticity) only when the in-situ disky component still dominated in mass. They 
concluded that gas accretion  
is needed to explain NSC formation in late-type spirals. 

\subsubsection{NGC\,2403}\label{subsub:ngc2403}
\begin{sidewaysfigure}[ht]
\centering
\resizebox{1.01\textwidth}{!}
{\includegraphics[scale=1.6, trim={1.3cm 0 0 0}]{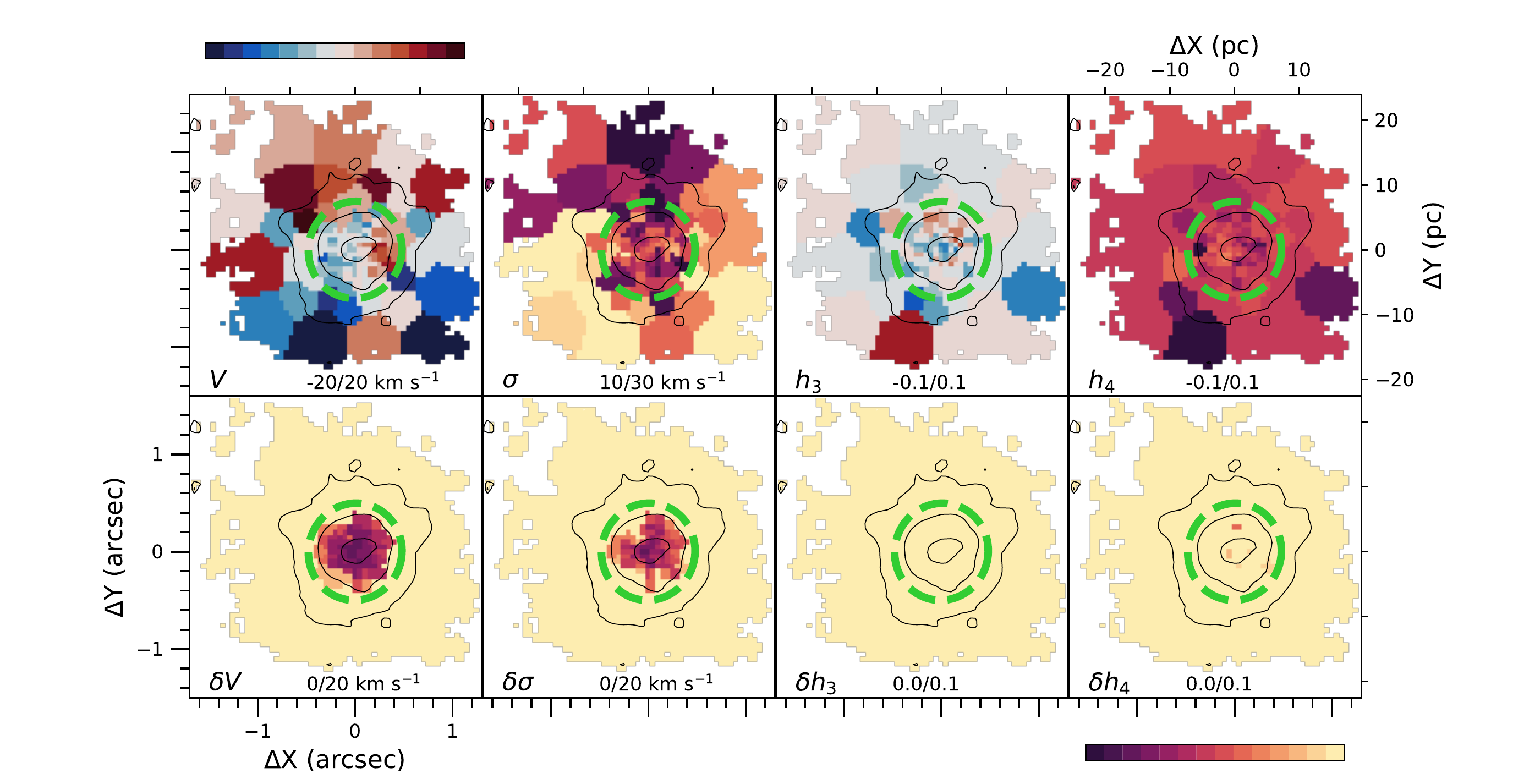}}
\caption{As Fig.\,\ref{fig:m32}, now for the nuclear region of  NGC\,2403. 
The original data cube was Voronoi-binned to a signal-to-noise ratio (S/N) of 15 per bin.
}
\label{fig:ngc2403}
\end{sidewaysfigure}
NGC\,2403 is very similar to M\,33 in its morphology, brightness, size, gas chemical properties and star-formation history  \citep{Garnett1997,Davidge2002,Davidge2003b}. 
The transient X-ray source detected by \citet{Yukita2007} suggests the presence of a low-mass BH, of similar properties to the one in M\,33's center. 
Being the second brightest galaxy of the M\,81 group, it is located in its outskirts \citep[e.g.,][]{Karachentsev2002,Williams2013,Carson2015}. Farther from M\,81 than M\,33 from M\,31 \citep[e.g.,][]{Davidge2003b}, its higher oxygen yield than M\,33 suggests it probably did not suffer from a similar gas stripping \citep{Garnett1997}. 
Nevertheless, it is approaching to M\,81 and has several faint dwarf satellites. A close one of them is old, metal poor and has no gas \citep{Karachentsev2002,Karachentsev2013,Carlin2016}.

NGC\,2403 has an extended and complex gas and dust structure \citep{Guelin1969,Shostak1973, Fraternali2002b,Bendo2007}. 
Apart from the warped HI disk, an additional component, moving towards the galactic center, was observed by \citet{Fraternali2001,Fraternali2002b}. It was interpreted as the result of a galactic fountain or an active gas channel between the disk and the halo \citep{Fraternali2002a}.
An additional cloud close to NGC\,2403, probably stripped from a satellite, was later discovered 
\citep{deBlok2014}.
The stellar component is distributed 
in a young undisturbed disk, 
and a more extended, thicker and fainter component, which is older and metal poor 
\citep{Davidge2003b,Barker2012,Williams2013}. 

Slow rotation is seen in the central region of this low-surface-brightness nucleus that appears to be almost face on (Fig.\,\ref{fig:ngc2403}). 
Its low velocity dispersion (slightly higher in the very center) is in agreement with a low-mass BH. 
This NSC is younger than the surrounding central region of the galaxy, as well as the one in M\,33, but even bluer 
\citep{Davidge2002}. However, the absence of significant emission from ionized gas suggests that the star formation might have recently halted \citep{Drissen1999,Boeker1999}. The NSC displays a larger size with longer wavelengths \citep{Carson2015}, as well as M\,33, suggesting that it was formed in different star-formation episodes, from gas funnelled from the different structures observed around.

\subsubsection{NGC\,2976}\label{subsub:ngc2976}
This bulgeless galaxy is classified as an Sc peculiar and it is located in the core of the M\,81 Group \citep[e.g.,][]{Karachentsev2002}. This region is covered by a system of HI clouds, filaments and bridges connecting galaxies, including NGC\,2976, to M\,81 \citep{Appleton1981,Chynoweth2008}. 
Past interactions of NGC\,2976 within the group are further supported by other studies 
\citep{Drzazga2016,Carozzi1980,Adams2012}.
The stellar populations are distributed in a young inner disk and in an old outer component (disk or halo)  \citep{Bronkalla1990,Bronkalla1992}. 
Inner-disk star formation, distributed in a ring-like structure, 
as well as outside-in gas depletion in the outer component, were probably triggered by the group environment (see also \citealt{Williams2010a}). 
Gas would have been stripped from the halo and/or channelled towards the galaxy center.
Another consequence of this might be the formation of a weak (gas-rich) bar, connecting two strong H$\alpha$-emission spots on both sides of the galaxy and favouring a potential nuclear starburst forming the NSC 
\citep{Tacconi1990,Daigle2006,MenendezDelmestre2007,Spekkens2007,Grier2011,Adams2012,Valenzuela2014}.

\begin{sidewaysfigure}[ht]
\centering
\resizebox{1\textwidth}{!}
{\includegraphics[scale=1.6, trim={1.3cm 0 0 0}]{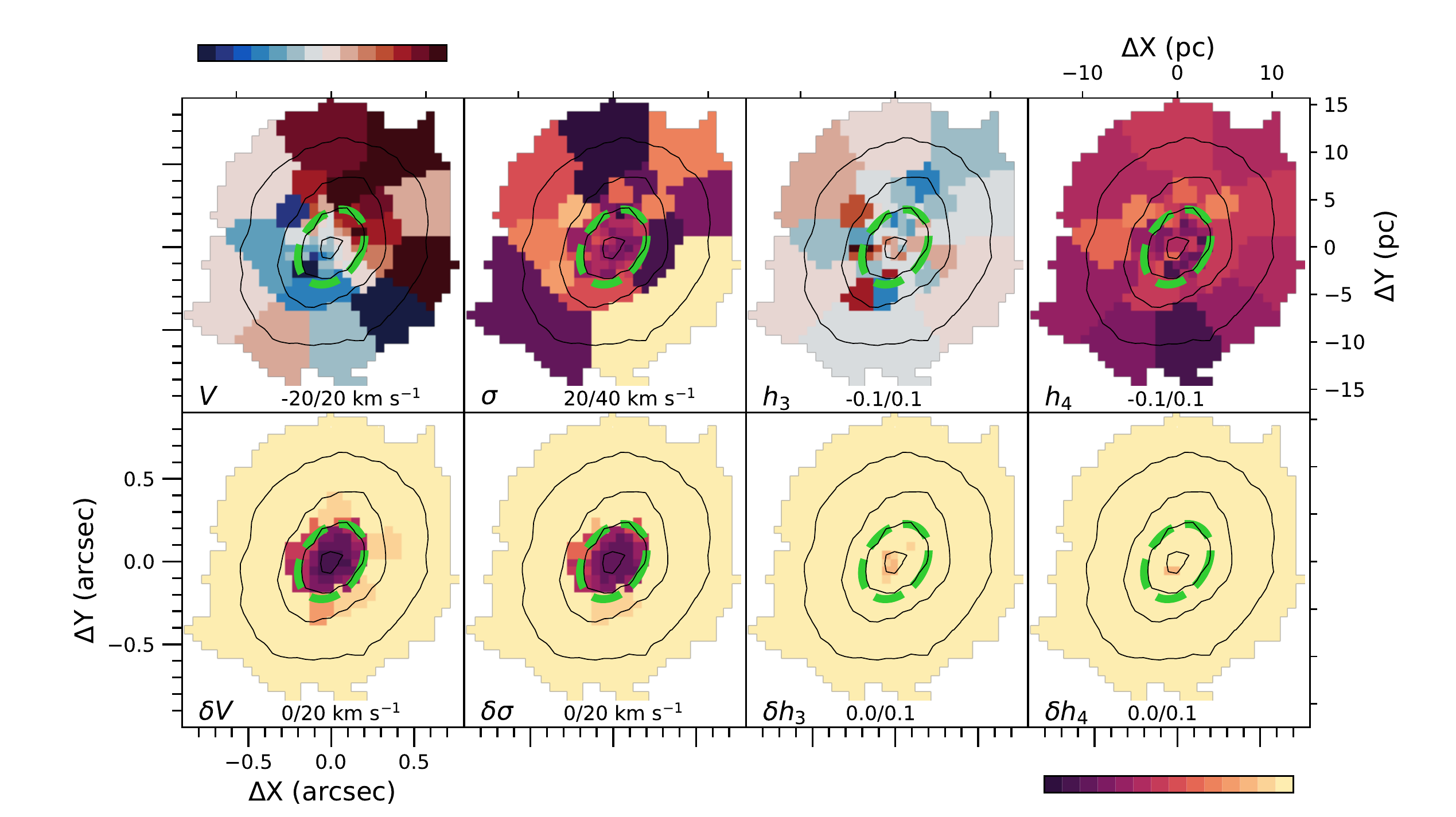}}
\caption{As Fig.\,\ref{fig:m32}, now for the nuclear region of NGC\,2976. 
The original data cube was Voronoi-binned to a signal-to-noise ratio (S/N) of 15 per bin.
}
\label{fig:ngc2976}
\end{sidewaysfigure}

The kinematics of this nucleus (Fig.\,\ref{fig:ngc2976}) shows clear rotation throughout the full FoV, even in the regions where large uncertainties warn us not to trust absolute values. Due to the low S/N, especially out of the \reff\,elliptical isophote, it is challenging to distinguish any structures in the maps of the higher moments $h_3$ and $h_4$. 
We know from photometry that this NSC displays some clumpiness 
and a significant flattening, as a disk oriented in the same way as the galaxy \citep{Carson2015}. 
The evidence provided for gas inflow towards the galactic center, with the bar playing a significant role, supports an in-situ formation for this NSC. 
However, it may have been formed in different star-formation episodes or by different formation mechanisms, since asymmetric stellar populations were identified by \citet{Carson2015}. 
It has a more compact bluer component to the north (with the HI structures in the north-east direction, e.g. \citealt{Chynoweth2008}) and a more extended and irregular redder component to the south. 

\subsubsection{NGC\,4244}\label{subsub:ngc4244}
NGC\,4244 is an edge-on late-type spiral, member of the M\,94 Group and probably weakly interacting with NGC\,4214 \citep[e.g.,][]{Seth2005a,Seth2005b,Comeron2011,Carson2015}.
Bulgeless, its brightest components are a prominent disk and a NSC. 
The disk was fitted vertically with two components, a thin and a thick disks, by \citet{Comeron2011}. It is characterized, at increasing distances from the midplane, respectively by young, intermediate and old stars \citep{Seth2005a,Seth2005b}. These populations break all at the same radius, pointing to a past interaction that may also explain the presence of a stellar diffuse component \citep{deJong2007,Seth2007}.
NGC\,4244 has a massive and thick HI disk. A warped and a flaring components, as well as other peculiar features in correspondence of star-forming regions, were observed \citep{Olling1996,Zschaechner2011}. At a smaller scale the stellar disk warps in the opposite direction to the HI disk, and shows signs of potential tidal interactions \citep{Comeron2011}. 
This galaxy may host in its center a massive BH of $\sim 10^5$\msun\,\citep{Hartmann2011,deLorenzi2013}. 

\begin{sidewaysfigure}[ht]
\centering
\resizebox{1.01\textwidth}{!}
{\includegraphics[scale=1.6, trim={1.3cm 0 0 0}]{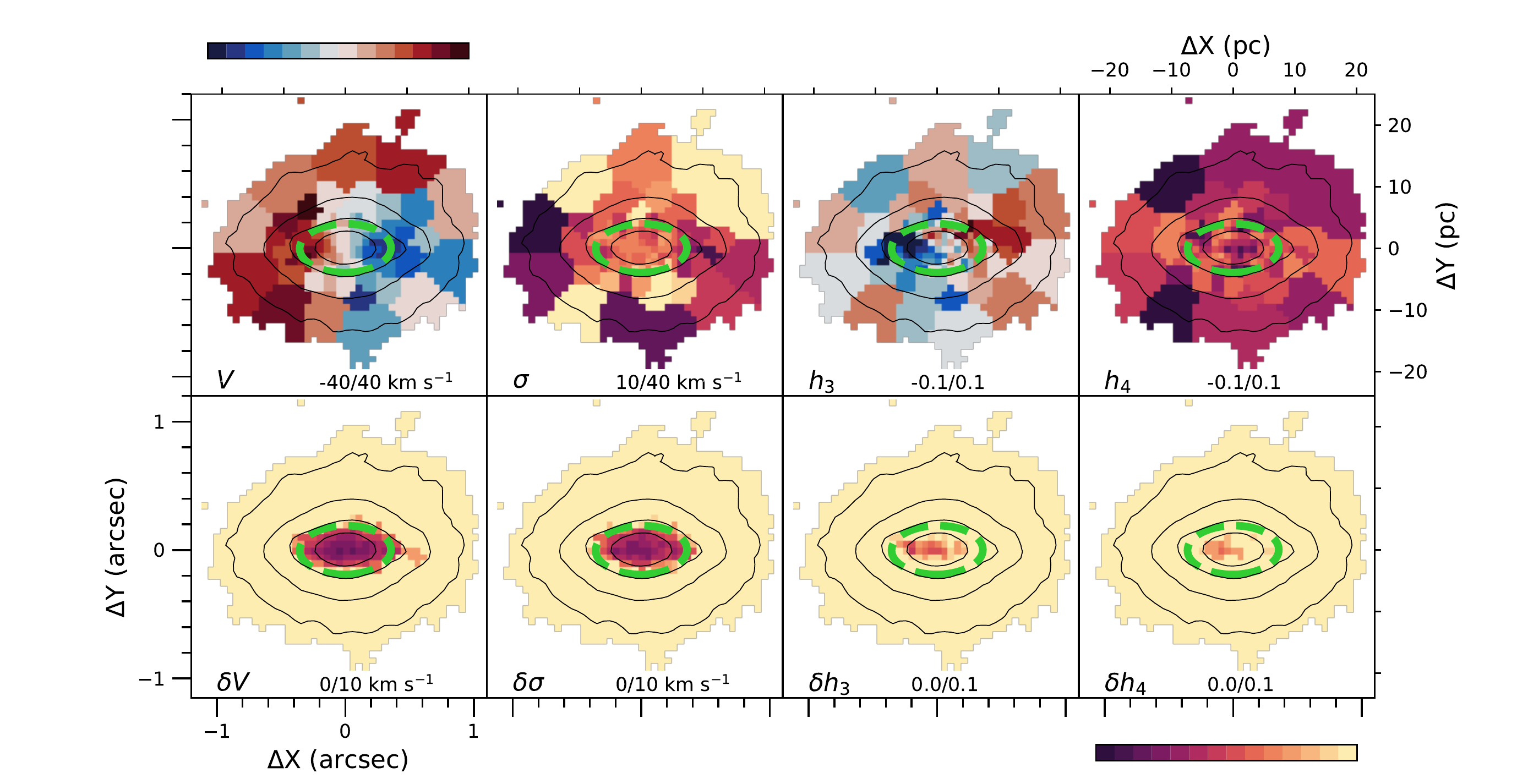}}
\caption{As Fig.\,\ref{fig:m32}, now for the nuclear region of NGC\,4244. 
The original data cube was Voronoi-binned to a signal-to-noise ratio (S/N) of 15 per bin.
}
\label{fig:ngc4244}
\end{sidewaysfigure}

The nearly edge-on nucleus of NGC\,4244 \citep{Hartmann2011} shows the highest ratio between circular and random motions in our sample (see Fig.\,\ref{fig:vsig_sample} and \ref{fig:lamRe_sample}), with the highest $\epsilon_{\rm{e}}$. 
Its kinematic maps are shown in Fig.\,\ref{fig:ngc4244} and are in agreement with the ones shown by \citet{Seth2008b}. 
Values of $\sigma$ within the central region, with reasonable uncertainties, are consistent with measurements from \citet{Ho2019}. 
A clear rotation pattern, anticorrelated with $h_3$, indicates a disky structure. 
In fact, the nucleus of NGC\,4244 is made up by a compact spheroidal component and a more extended disk, co-rotating with the HI disk \citep{Seth2008b,Carson2015}. 
Multiple stellar populations were identified in this NSC, with 
young stars ($\sim 100$\,Myr) dominating the disk component. 
Old and more metal-poor stellar populations dominate above and below the midplane \citep{Seth2006,Seth2008b}. 
A combined accretion of gas and star clusters from the galaxy disk was proposed for the formation of this massive NSC ($\sim 10^7$\msun), supported by results from simulations and dynamical models \citep{Hartmann2011,deLorenzi2013}. 

\subsubsection{NGC\,4449}\label{subsub:ngc4449}
NGC\,4449 is a luminous irregular galaxy with intense recent star formation 
and a rich population of star clusters of all ages \citep{Kumari2017,Whitmore2020}. 
Numerous studies presented hints of recent interactions, such as stellar tidal streams and, on the west side of the galaxy, a star cluster with a tidal structure which could be the remnant nucleus of a disrupted dwarf galaxy \citep{Martinez2012,Annibali2008}.
Although NGC\,4449 appears relatively isolated, since its only close companion (in the projected space) is DDO\,125, a past interaction between the two galaxies was suggested, e.g., by \citet{Theis2001} and \citet{valdez2002}. This would explain the complex morphology and kinematics of the extended gas structure around NGC\,4449 \citep{Bajaja1994,Hunter1997a,Hunter1997b}. 
Pointing to the same direction, counter-rotation has been observed in the inner gas component (within $\sim4$\,kpc) with respect to the outer gas envelope \citep{Hunter1998}.

\begin{sidewaysfigure}[ht]
\centering
\resizebox{1.01\textwidth}{!}
{\includegraphics[scale=1.6, trim={1.3cm 0 0 0}]{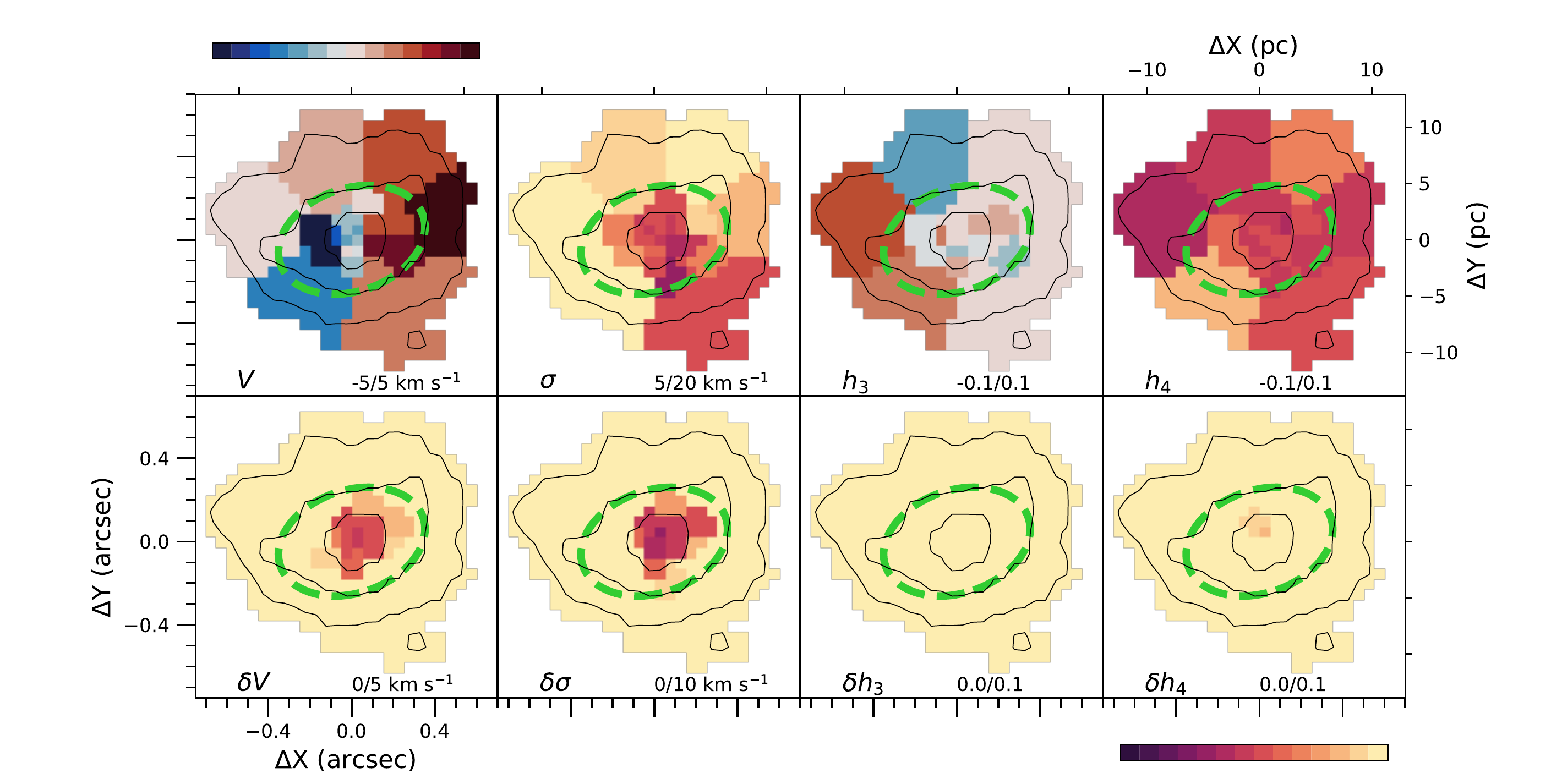}}
\caption{As Fig.\,\ref{fig:m32}, now for the nuclear region of NGC\,4449. 
The original data cube was Voronoi-binned to a signal-to-noise ratio (S/N) of 15 per bin.
}
\label{fig:ngc4449}
\end{sidewaysfigure}

In Fig.\,\ref{fig:ngc4449}, we observe slow rotation and low values of $\sigma$. 
The kinematic analysis of this nucleus was the most challenging in our sample, due to the combination of the lowest S/N among our data cubes and the low absolute values of the kinematic parameters (proving the limits of our method). 
The nucleus of NGC\,4449, associated with intense emission of ionized gas, shows no evidence of old populations and hosts the youngest stars of the galaxy ($\sim10$\,Myr old) 
\citep{Boeker1999,Gelatt2001,Annibali2008}. 
The latter indicate a very recent starburst, probably
a consequence of a violent event that doubled the star-formation rate between 10 and 16\,Myr ago \citep{Cignoni2018,Whitmore2020}. 

The relatively low metallicity shows that these stars were formed from gas that was not significantly enriched \citep{Boeker2001}.
This is also confirmed by the lower metallicity of ionized gas in the central region than in the rest, suggesting metal-poor gas accretion during a potential recent merger \citep{Kumari2017}. 
This gas might have been funnelled towards the galactic center by the S-shaped structure that was observed in this nuclear region \citep{Gelatt2001}. 
This might be a bar-like debris of a past interaction, probably a small (accreted) spiral seen edge-on. 
This structure (9.5\,pc long, \citealt{Gelatt2001}), might be the rotating structure in the center of our FoV. 

Looking like a good candidate for a pure gas-accretion NSC formation, we would have expected to find this nucleus in an upper location in Fig.\,\ref{fig:vsig_sample} and \ref{fig:lamRe_sample}. 
Its peculiar nature (and of its host), and the potential composition of different structures in the LOS as a result of mergers, might explain why it is the lowest point for late-type galaxies in our diagrams.
In alternative, its very young age and mass lower than $10^7$\msun\,\citep{Georgiev2016} might be implying that it did not have time to grow enough and reach the high rotation levels typical of the most massive NSCs such as M\,32, NGC\,4244, NGC\,5102 and NGC\,7793.

\begin{sidewaysfigure}[ht]
\centering
\resizebox{1.01\textwidth}{!}
{\includegraphics[scale=1.6, trim={1.3cm 0 0 0}]{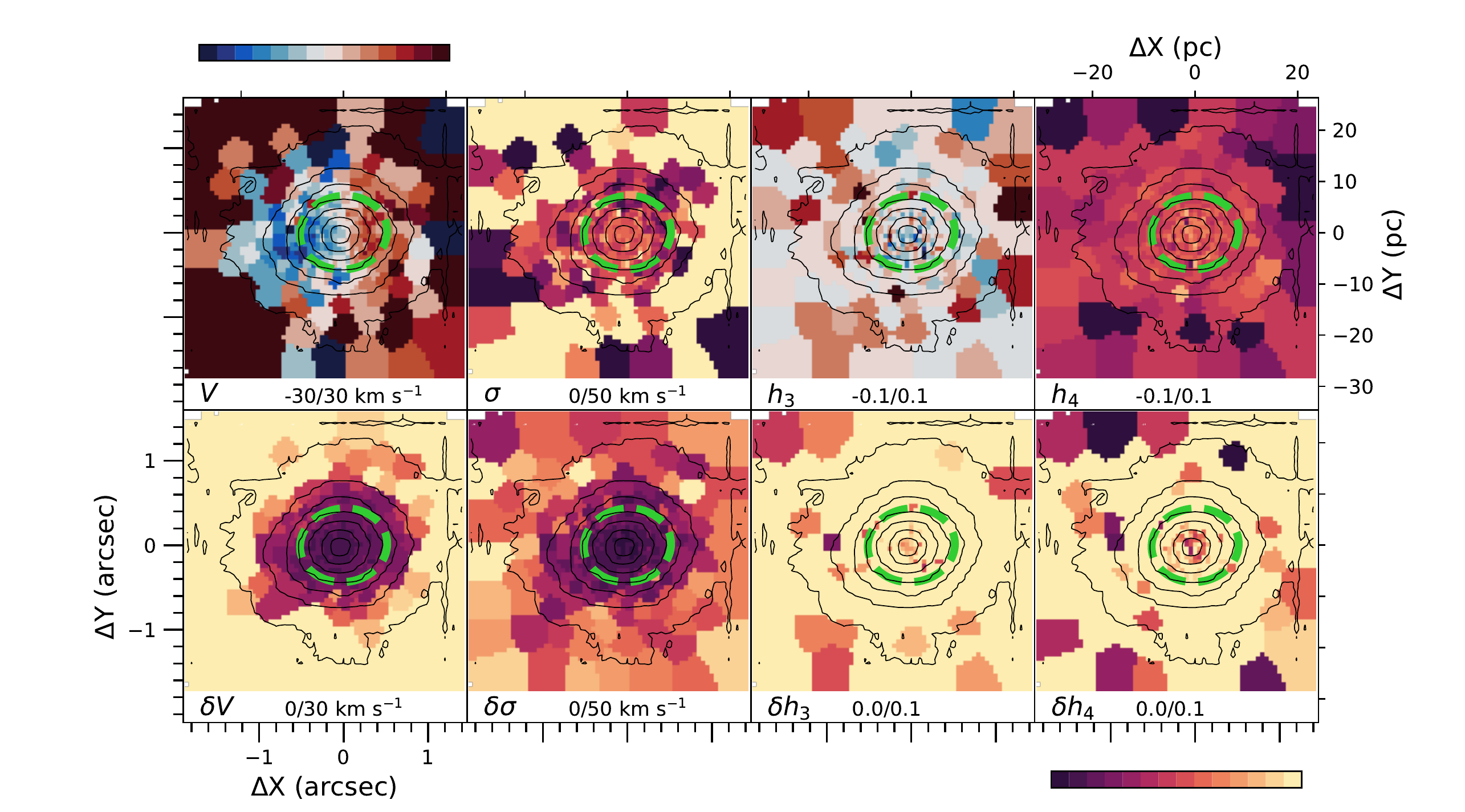}}
\caption{As Fig.\,\ref{fig:m32}, now for the nuclear region of NGC\,7793. 
The original data cube was Voronoi-binned to a signal-to-noise ratio (S/N) of 25 per bin.
}
\label{fig:ngc7793}
\end{sidewaysfigure}

\subsubsection{NGC\,7793}\label{subsub:ngc7793}
NGC\,7793 is a bright spiral galaxy, member of the Sculptor group. It has two dwarf close companions in the same subgroup \citep{Karachentsev2003}.
Considered as the prototype of Sd galaxies \citep{Shapley1943}, it has no bulge (but a bright nucleus). It is rich in star-forming regions in its multiple fragmented and clumpy spiral arms 
\citep{deVaucouleurs1980,Smith1984,Sacchi2019}. 
Both the extended HI and diffuse H$\alpha$ components 
show a declining rotation curve, 
some level of non-circular motions, and a warp in the outskirts. \citep{Dicaire2008,Carignan1990,Davoust1980,deVaucouleurs1980}.
The stellar disk of NGC\,7793 is made up by an old underlying component and young populations in the spiral arms \citep{deVaucouleurs1980,Davidge1998}.
Old stars, with an upturn in their brigthtness profile in the outskirts, extend farther than the young populations and than the HI disk. Radial migration was proposed by \citet{Radburn-Smith2012} to explain this. The resolved star-formation history of NGC\,7793 reveals an increase in the star-formation rate in time and spatially from the inner to the outer regions, pointing to a inside-out growth of the disk \citep{Sacchi2019}. 
A massive BH of $\lesssim 5\times 10^5$ was recently estimated by Neumayer et al. (in prep.).

The NSC in NGC\,7793 is rather massive ($7.8\times 10^7$\msun, \citealt{Walcher2005}). 
It shows clear rotation around its minor axis, with a slightly asymmetric pattern (Fig.\,\ref{fig:ngc7793}). 
Velocity dispersion assumes slightly higher values in the very central region, that are in agreement with the average velocity dispersion from \citet{Walcher2005} ($\sim 24.6$\kms), measured over a 10-arcsec-long slit. 
It is difficult to distinguish any structures in the $h_3$ and $h_4$ maps, with high relative uncertainties. 
NGC\,7793's nucleus, dominated by very young stars, is one of the many star-formation regions in this galaxy 
\citep{Diaz1982, Shields1992, Walcher2006, Kacharov2018}. 
A color gradient and a UV-prominent ring structure showed that youngest stars are mainly located in the outer part of the NSC, suggesting circumnuclear star formation \citep{Carson2015}. 
The position of this nucleus in the \vsRe\,and \lamRe\,diagrams (Fig.\,\ref{fig:vsig_comp} and \ref{fig:lamRe_comp}), in the upper region, suggests in-situ formation as the dominant formation mechanism. 
On the other hand, results on the stellar-population properties of this NSC, consisting in a variety of ages and metallicities, point towards multiple star clusters mergers contributing the older components \citep{Kacharov2018}. 

\newpage
\section{$\abs{V}/\sigma$ radial distribution in the nuclear regions of the eleven early and late-type galaxies in our sample} \label{app:Vsigma_rad}
\begin{figure}
\centering
\scalebox{0.64}
{\includegraphics[scale=1]{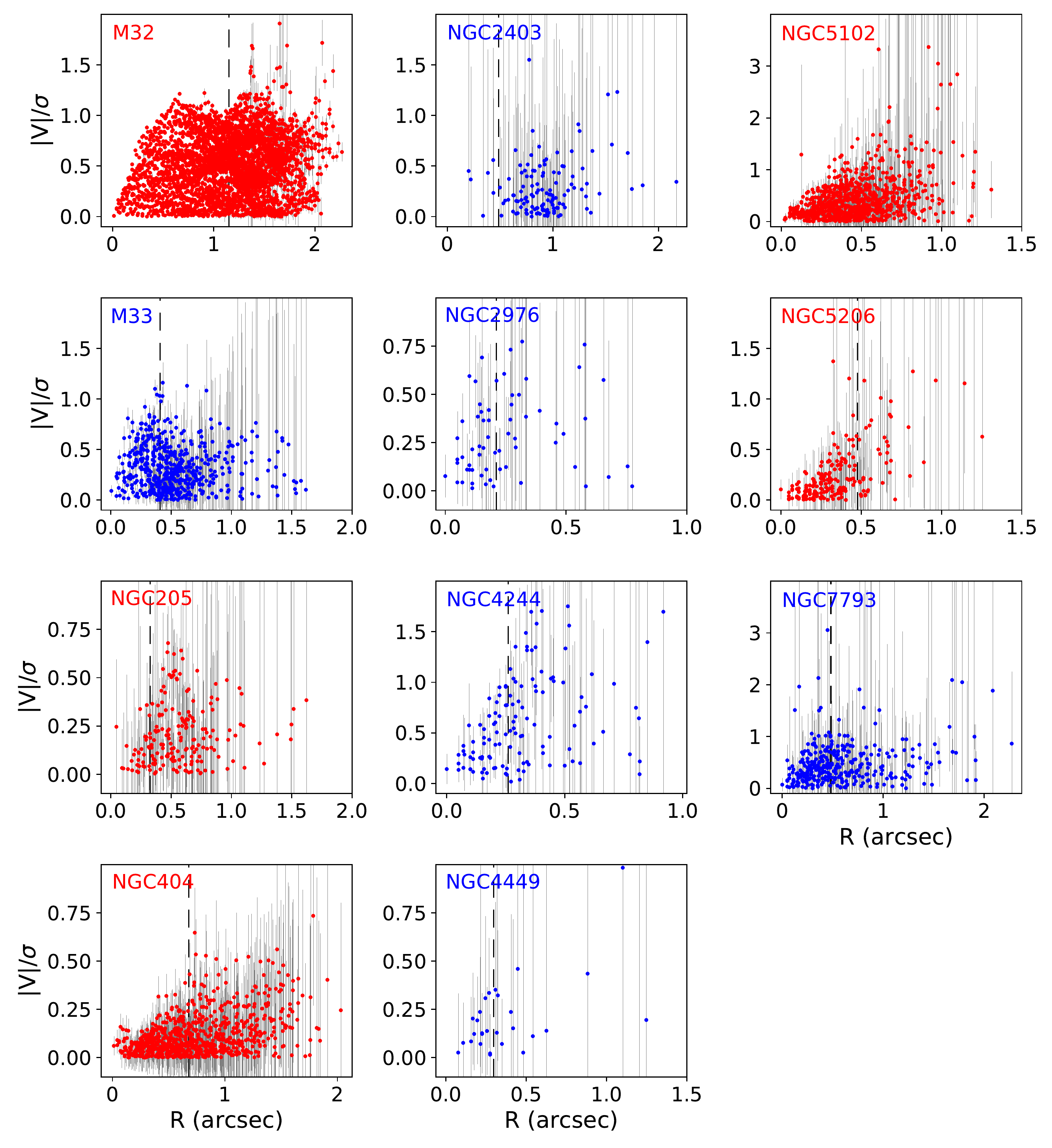}} 
\caption{Radial distributions of $\abs{V}/\sigma$ in the nuclear regions of the eleven galaxies in our sample, each one in one panel. Each point corresponds to a Voronoi bin. Points in early-type galaxies are plotted in red and the ones in late-type galaxies in blue. Galaxy names are indicated on top left of each panel with the same color scheme. Error bars are plotted in grey. 
The horizontal axes correspond to the projected radius in the FoV, as it is observed in our NIFS and SINFONI data. \reff\,is indicated by a vertical dashed line in each panel. 
}
\label{fig:vsig_prof}
\end{figure}

We show in Fig.\,\ref{fig:vsig_prof} the radial profiles of $\abs{V}/\sigma$ for each nucleus in our sample. Each point corresponds to $V$ and $\sigma$ in each Voronoi bin, as mapped in figures from \ref{fig:m32} to \ref{fig:ngc7793}. 
Points and names of early-type galaxies are plotted in red and the ones of late-type galaxies in blue. Error bars, plotted in grey, correspond to the uncertainties mapped in figures from \ref{fig:m32} to \ref{fig:ngc7793}. 
These plots may help the assessment and interpretation of Fig.\,\ref{fig:vsig_sample}, \ref{fig:lamRe_sample}, \ref{fig:vsig_comp}, and \ref{fig:lamRe_comp}. 

\section{Measurements of $\epsilon_{\rm{\lowercase{e}}}$, \vsRe\,and \lamRe\,for our sample of eleven nuclei} \label{app:meas}
We provide here our results for $\epsilon_{\rm{e}}$, \vsRe\,and \lamRe, calculated as explained in Sect.\,\ref{subsec:vsigma}, for the eleven nuclei in our sample. These values are plotted in Fig.\,\ref{fig:vsig_sample}, \ref{fig:lamRe_sample}, \ref{fig:vsig_comp}, and \ref{fig:lamRe_comp}. 
Values corrected from the PSF effect were calculated as explained in Appendix\,\ref{app:PSF} and are shown in Fig.\,\ref{fig:vsig_corr} and \ref{fig:lamRe_corr}. $\left(V/\sigma \right)_\text{e}^{\rm{corr}}$ and ${\lambda_R}_\text{e}^{\rm{corr}}$ for NGC\,4244, NGC\,5102 and NGC\,5206 are upper limits. 

\begin{table*}[!ht]
\centering
\caption{Kinematic properties of our sample of eleven galactic nuclei. Columns from left to right: galaxy name, ellipticity, ellipticity error, \vsRe, its uncertainty, \lamRe, its uncertainty, PSF corrected values of \vsRe\,and PSF corrected \lamRe. 
All values and their uncertainties were calculated as explained in Sect.\,\ref{subsec:vsigma} and corrected from the PSF effects as explained in Appendix\,\ref{app:PSF}.}
\centering
\begin{tabular}{l|c|c|c|c|c|c|c|c} 
\hline\hline
\footnotesize \centering
Galaxy name & $\epsilon_{\rm{e}}$ & $\delta \epsilon_{\rm{e}}$& \vsRe &$\delta$\vsRe & \lamRe &  $\delta$\lamRe &$\left(V/\sigma \right)_\text{e}^{\rm{corr}}$&${\lambda_R}_\text{e}^{\rm{corr}}$ \\  
\hline\hline
M\,32      & 0.187 & 0.003 & 0.504 & 0.001 & 0.456 & 0.001 & 0.589 & 0.526 \\
M\,33      & 0.113 & 0.024 & 0.419 & 0.007 & 0.354 & 0.006 & 0.608 & 0.529 \\
NGC\,205   & 0.164 & 0.034 & 0.217 & 0.013 & 0.169 & 0.012 & 0.405 & 0.296 \\
NGC\,404   & 0.091 & 0.009 & 0.089 & 0.003 & 0.079 & 0.003 & 0.122 & 0.108 \\
NGC\,2403 & 0.042 & 0.018 & 0.294  & 0.041 & 0.254 & 0.020 & 0.420 & 0.362 \\
NGC\,2976 & 0.287 & 0.063 & 0.310  & 0.038 & 0.273 & 0.026 & 0.616 & 0.594 \\
NGC\,4244 & 0.465 & 0.035 & 0.572  & 0.018 & 0.499 & 0.014 & $<$0.858 & $<$0.677 \\
NGC\,4449 & 0.318 & 0.034 & 0.203  & 0.081 & 0.165 & 0.033 & 0.357 & 0.275 \\
NGC\,5102 & 0.243 & 0.004 & 0.325  & 0.007 & 0.351 & 0.005 & $<$0.440 & $<$0.465 \\
NGC\,5206 & 0.079 & 0.020 & 0.219  & 0.012 & 0.195 & 0.010 & $<$0.274 & $<$0.243 \\
NGC\,7793 & 0.157  & 0.015 & 0.392 & 0.012 & 0.367 & 0.010 & 0.528 & 0.494 \\
\hline
\hline
\end{tabular}
\label{tab:meas}
\end{table*}


\section{PSF measurements and their impact on the observed \vsRe\,and \lamRe}\label{app:PSF}
We follow the approach of \citet{Harborne2020} to assess the impact of atmospheric seeing on our observational measurements of the nuclear \vsRe\,and \lamRe. 
They proposed general analytic corrections, modelled making use of mock observations of numerical simulations, accounting for the PSF-effect on \vsRe\,and \lamRe\,on galaxies of different Hubble types. 
From their Eq.\,16, 17 and 18, the corrected \vsRe\,and \lamRe\,are:
\begin{equation}
\left(\frac{V}{\sigma}\right)_{\rm{e}}^{\rm{corr}} = 10^{\left[ \log_{10}\left(\frac{V}{\sigma}\right)_{\rm{e}} - \Delta \left(\frac{V}{\sigma}\right)_{\rm{e}}^{\rm{corr}} \right]}
\end{equation}
and
\begin{equation}
{\lambda_R}_\text{e}^{\rm{corr}} = 10^{\left[ \log_{10}\left( {\lambda_R}_\text{e}\right) - \Delta {\lambda_R}_\text{e}^{\rm{corr}} \right]}
\end{equation}
where:
\begin{equation}
\Delta \left(\frac{V}{\sigma}\right)_{\rm{e}}^{\rm{corr}} = \left\lbrace \frac{7.55}{1+\exp\left[ 4.42 \left(\frac{\sigma_{\text{PSF}}}{R_{\text{e}}^{\text{maj}}}\right)^{1.55} + 2.73\right]} - 0.46 \right\rbrace + 3 \left(\frac{\sigma_{\text{PSF}}}{R_{\text{e}}^{\text{maj}}}\right)\times
\left[ -0.10 \epsilon_{\rm{e}} + 0.024\log_{10}\left(n_{\rm{NSC}}\right) +0.064\right]
\end{equation}
and
\begin{equation}
\Delta {\lambda_R}_\text{e}^{\rm{corr}} = 
\left\lbrace \frac{7.48}{1+\exp\left[ 4.08 \left(\frac{\sigma_{\text{PSF}}}{R_{\text{e}}^{\text{maj}}}\right)^{1.60} + 2.89\right]} - 0.39 \right\rbrace + \left(\frac{\sigma_{\text{PSF}}}{R_{\text{e}}^{\text{maj}}}\right) \times
\left[ 0.10 \epsilon_{\rm{e}} -0.22\log_{10}\left(n_{\rm{NSC}}\right) +0.10\right]
\end{equation}

\begin{table*}[!ht]
\centering
\caption{PSF measurements for our observations (all galaxies in the sample, except NGC\,4449). Two fitting components were used for most galaxies. $<\text{FWHM}_\text{PSF}>$ in the second-to-last column is the light weighted average of the $\text{FWHM}_\text{PSF}$ of the different profiles.
}
\centering
\begin{tabular}{l|c|c|c|c|r} 
\hline\hline
\footnotesize
Galaxy & component & $\text{FWHM}_\text{PSF}$ & light fraction & $<\text{FWHM}_\text{PSF}>$&\multirow{2}{*}{Reference}\\ 
name & of the fit & (arcsec) & \% & (arcsec) & \\ 
\hline\hline
\multirow{2}{*}{M\,32}          &  Gaussian & 0.23 & 45 &\multirow{2}{*}{0.46} &\multirow{2}{*}{\citet{Seth2010a}}\\
&  Moffat & 0.66$^{(1)}$ & 55 &\\
\hline
\multirow{2}{*}{M\,33}          &  Gaussian & 0.09 & 65 &\multirow{2}{*}{0.31} &\multirow{2}{*}{this work}\\
&  Gaussian & 0.71 & 35 &\\
\hline
\multirow{2}{*}{NGC\,205}         &  Gaussian & 0.25 &68&\multirow{2}{*}{0.40}&\multirow{2}{*}{\citet{Nguyen2018}}\\
&  Moffat & 0.73$^{(1)}$&32&\\
\hline
\multirow{2}{*}{NGC\,2403}          &  Gaussian & 0.22 & 86 &\multirow{2}{*}{0.38} &\multirow{2}{*}{this work}\\
&  Gaussian & 1.35 & 14 &\\
\hline
\multirow{2}{*}{NGC\,2976}          &  Gaussian & 0.13 & 68 &\multirow{2}{*}{0.34} &\multirow{2}{*}{this work}\\
&  Gaussian & 0.78 & 32 &\\
\hline
\multirow{2}{*}{NGC\,404}   &  Gaussian & 0.12&50&\multirow{2}{*}{0.44}&\multirow{2}{*}{\citet{Seth2010b}}\\
&  Moffat &0.75$^{(1)}$ &50 &\\
\hline
\multirow{1}{*}{NGC\,4244} &  Gaussian & 0.23 & 100&\multirow{1}{*}{0.23} &\multirow{1}{*}{\citet{Seth2008b}}\\
\hline
\multirow{2}{*}{NGC\,5102}  &  Gaussian & 0.08&35&\multirow{2}{*}{0.56}&\multirow{2}{*}{\citet{Nguyen2018}}\\
&  Gaussian & 0.82&65&\\
\hline
\multirow{2}{*}{NGC\,5206}  &  Gaussian & 0.12&60&\multirow{2}{*}{0.24}&\multirow{2}{*}{\citet{Nguyen2018}}\\
&  Gaussian &0.42 &40&\\
\hline
\multirow{2}{*}{NGC\,7793}  &  Gaussian & 0.17 &67&\multirow{2}{*}{0.30}&\multirow{2}{*}{this work}\\
&  Gaussian & 0.57&33&\\
\hline
\hline
\end{tabular}
\label{tab:psf}
\begin{tablenotes}
\item {\footnotesize Notes. (1) For Moffat distributions described as $\Sigma (r)=\Sigma_0 \left[ 1 + \left(r/r_d \right)^2 \right]^{-4.765}$ \citep{Seth2010a,Seth2010b,Nguyen2018}, $\text{FWHM}_\text{PSF} = 2r_d \sqrt{2^{1/4.765}-1}$ \citep{Trujillo2001}.
}
\end{tablenotes}
\end{table*}
These corrections are functions of the ratio $\sigma_{\text{PSF}}/R_{\text{e}}^{\text{maj}}$, between the standard deviation of the PSF distribution $\sigma_{\text{PSF}}=\text{FWHM}_\text{PSF}/(2\sqrt{2\ln 2})$, and the major axis of the NSC half-light ellipse $R_{\text{e}}^{\text{maj}}$. 
In addition, the impact of the PSF depends on the concentration of the light, indicated by the S\'ersic index $n_{\rm{NSC}}$. 

We show in Table\,\ref{tab:psf} the details of the available PSF measurements, mentioned in Sect.\,\ref{sec:obs}, for our observations. 
For galaxies in Table\,\ref{tab:psf}, we used the light-weighted average FWHM of the different PSF-fitting components ($<\text{FWHM}_\text{PSF}>$). 
For NGC\,4449, since it was not possible to derive a PSF for our data cubes from optical images (due to the complexity of its central region), we used the average PSF size measured for our observations with the same instrument and mode (NIFS with LGS AO, see Sect.\,\ref{sub:nifs}). This corresponds to $\text{FWHM}_\text{PSF}=0.345$\,arcsec. 
As for $n_{\rm{NSC}}$,  we used the values in Table\,\ref{tab:sample}. 
When two values of $n_{\rm{NSC}}$ are given in the table, because the NSC was fitted with a double S\'ersic profile, we adopted the highest index to obtain the largest correction and being more conservative. Therefore we obtained upper limits for the corrected values of NGC\,4244, NGC\,5102 and NGC\,5206. 
Since for NGC\,4449 we did not find any available value for $n_{\rm{NSC}}$, we approximate it to the mean S\'ersic index of the rest of the sample ($n_{\rm{NSC}}\sim2.9$). 

\begin{figure}
\centering
\scalebox{0.42}
{\includegraphics[scale=1,page=3,trim={0.5cm 0.7cm 0.5cm 0.7cm}]{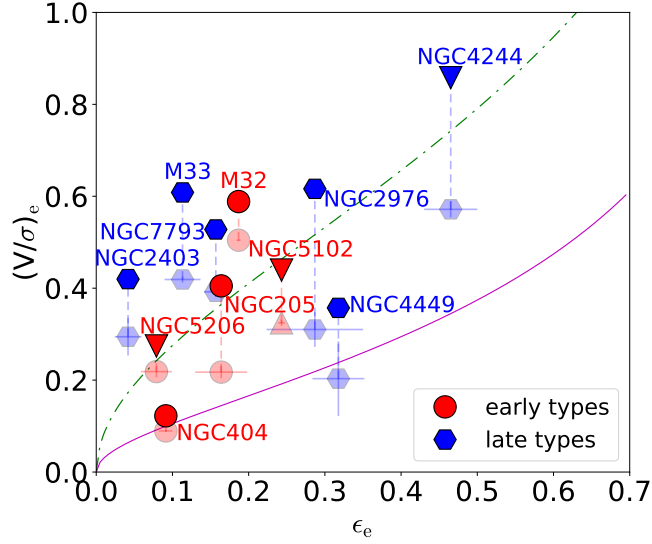}}
\caption{Estimate of the PSF impact on the \vsRe$ - \epsilon_{\rm{e}}$ diagram for the sample of eleven nuclei studied in this work. 
Shaded points indicate the observed \vsRe\,(as in Fig.\,\ref{fig:vsig_sample}) while filled points are the estimated values of \vsRe\,after PSF correction. 
The names of the host galaxies are indicated close to the individual corrected points. 
Nuclei of early-type galaxies are plotted in red (circles or downwards triangles) while the ones in late types are plotted in blue (hexagons or a downwards triangle). 
Downwards triangles indicate upper limits. 
The green dash-dotted and the magenta solid lines are as in Fig.\,\ref{fig:vsig_sample}. 
}
\label{fig:vsig_corr}
\end{figure}

\begin{figure}
\centering
\scalebox{0.42}
{\includegraphics[scale=1,page=4,trim={0.5cm 0.7cm 0.5cm 0.7cm}]{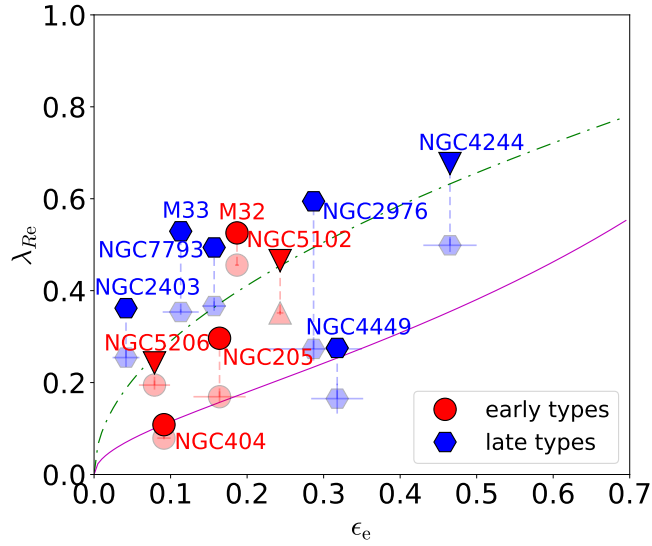}}
\caption{Estimate of the PSF impact on the \lamRe$ - \epsilon_{\rm{e}}$ diagram for the sample of eleven nuclei studied in this work. Symbols, lines and colors are as in Fig.\,\ref{fig:vsig_corr}. 
}
\label{fig:lamRe_corr}
\end{figure}

The obtained values of $\left(\frac{V}{\sigma}\right)_{\rm{e}}^{\rm{corr}}$ and ${\lambda_R}_\text{e}^{\rm{corr}}$ are indicated in Table\,\ref{tab:meas} (Appendix\,\ref{app:meas}). 
They are shown in the $\left(\frac{V}{\sigma}\right)_{\rm{e}}^{\rm{corr}} - \epsilon_{\rm{e}}$ and ${\lambda_R}_\text{e}^{\rm{corr}} - \epsilon_{\rm{e}}$ diagrams plotted in Fig.\,\ref{fig:vsig_corr} and \ref{fig:lamRe_corr}. 
Nuclei hosted by early and late-type galaxies are indicated respectively with red circles and blue hexagons. Upper limits are indicated with downwards triangles. 
In the figure, the corrected points are compared to the uncorrected ones from Fig.\,\ref{fig:vsig_sample} and \ref{fig:lamRe_sample} (shaded symbols). It is shown how they move upwards when the effect of the PSF is taken into account. 
However, each one of them does it in a different amount. 
While the PSF impact is relatively small for the nuclei in NGC\,404 and NGC\,5206, 
it is not so small for NGC\,205. Its PSF correction, the largest among the early-type galaxies in our sample and mainly due to the small size of the NSC, drives the red circle to \vsRe\,values not so different from NGC\,2403 and NGC\,5102. 
Except for NGC\,205, PSF corrections are in general more significant for late-type galaxies. 
This is due to their stronger rotation, in some cases combined with a larger concentration of the nucleus in the center. In fact, the PSF impact is more significant for nuclei with higher $n_{\rm{NSC}}$ and $\epsilon_{\rm{e}}$ and with smaller \reff\,relative to the PSF size. In particular, the highest $n_{\rm{NSC}}$ determines a strong PSF effect for NGC\,2976's nucleus.

\end{document}